\documentclass[twocolumn,amsmath,amssymb,aps,longbibliography]{revtex4-2}
\usepackage{graphicx}
\usepackage{upgreek}
\usepackage{color}
\usepackage{bibunits}
\usepackage{natbib}
\usepackage{xcolor}
\usepackage{comment}
\usepackage[colorlinks=true, linkbordercolor=red, citebordercolor=green, urlbordercolor=cyan, pdfborderstyle={/S/U/W 1}]{hyperref}

\usepackage{soul}
\begin{document}

\title{Single-mode Quantum Non-Gaussian Light from Warm Atoms}
\author{Jarom\'ir Mika,  Luk\'a\v s Lachman, Tom\'a\v s Lamich, Radim Filip, Luk\'a\v s Slodi\v cka}

\affiliation{Department of Optics, Palack\'{y} University, 17. listopadu 12, 771 46 Olomouc, Czech Republic}

\date{\today}

\begin{abstract}
The distributed quantum information processing and hybridization of quantum platforms raises increasing demands on the quality of light-matter interaction and realization of efficient quantum interfaces. This becomes particularly challenging for needed states possessing fundamental quantum non-Gaussian (QNG) aspects. They correspond to paramount resources in most potent applications of quantum technologies. We demonstrate the generation of light with provably QNG features from a tunable warm atomic ensemble in a single-mode regime. The light is generated in a spontaneous four-wave mixing process in the presence of decoherence effects caused by a large atomic thermal motion. Despite its high sensitivity to any excess noise, a direct observability of heralded QNG light could be achieved due to a combination of a fast resonant excitation, large spectral bandwidth, and a low absorption loss of resonant photons guaranteed by the source geometry.
\end{abstract}


\maketitle

\section*{Introduction}

Nonclassical states of light have provided a paramount resource for pioneering studies of quantum aspects of radiation and demonstrations of quantum paradigmatic phenomena in matter. They allowed a successful initial tests of quantum information theory and evolved into a natural carrier of quantum information over long distances. In modern applications of quantum optics, they have developed into an efficient interconnect between various quantum platforms with complementary properties~\cite{kurizki2015quantum} and correspond to an indispensable resource for quantum communication tasks. The generation of nonclassical single photons has since become nearly a rudimentary experimental exercise and it has been achieved with provably strong sub-Poissonian character in a wide range of implementations~\cite{slussarenko2019photonic}.

However, for the majority of the prospective applications of quantum technologies, a mere nonclassicality cannot provide a boosting resource over the analogous classical approach, but the enhancement must be supplied by the sensitive quantum non-Gaussian (QNG) properties available either as QNG input states or through the controllable nonlinear QNG interactions~\cite{lvovsky2020production}. By the definition, QNG states hallmark intrinsically nonlinear character of the source~\cite{hudson1974wigner,walschaers2021non} and already serve as an indispensable resources for the nontrivial character of quantum sensing~\cite{mccormick2019quantum,wolf2019motional}, and error correction~\cite{michael2016new,hu2019quantum,campagne2020quantum,fluhmann2019encoding}, dominantly for motional states of trapped ions and microwave radiation in the superconducting circuits. Despite of the optical implementations being significantly more challenging, past two decades brought several stimulating proof-of-principle demonstrations of the QNG states from the heralded cavity OPO sources~\cite{lvovsky2001quantum, zavatta2004tomographic,ourjoumtsev2006generating,ourjoumtsev2006quantum,miwa2014exploring,bimbard2014homodyne,le2018slowing,zapletal2021experimental,ra2020non}, and recently also from heralded single-atom source~\cite{hacker2019deterministic}, signified by the observation of the negativity of Wigner function~\cite{schleich2011quantum}. The analysis of optical fields experiencing strong attenuation, where evaluations based on the negativity of Wigner function fail, has been enabled by the development and implementation of the QNG criteria based on the precise knowledge of the few lowest photon-number probabilities~\cite{straka2014quantum,jevzek2011experimental,higginbottom2016pure,jevzek2012experimental}. Such states of light can already find applications in quantum communication and interfaces~\cite{lasota2017sufficiency,rakhubovsky2017photon}. Although the provable QNG properties of single-photon states have been unambiguously demonstrated in several optical platforms with complementary features, the crucial ability to efficiently interact with matter and form together complex interfering quantum systems remains elusive~\cite{lvovsky2020production,niset2009no,eisert2002distilling,ghose2007non,ohliger2010limitations}. Importantly, utilization of QNG light in the interaction with matter requires a well defined degrees of freedom corresponding to ideally a single optical mode, which would enable an efficient excitation of atomic or solid-state transitions.

The generation of QNG light source capable of coherent and efficient interaction with atoms and with other QNG light sources has been approached using various experimental platforms. Intrinsically single-photon emitters, including single trapped cold atoms or ions, allow for a direct observability of genuine QNG properties due to a good isolation from an environment and are working in a provably close-to a single mode regime, however, typically provide either low photon rates~\cite{hacker2019deterministic} or low photon detection efficiencies~\cite{higginbottom2016pure}. QNG sources implemented by a heralded generation of states approaching single-photon Fock states in various optical nonlinear processes~\cite{jevzek2011experimental,straka2014quantum,lachman2019faithful,lvovsky2020production,ourjoumtsev2006generating,ra2020non,sychev2017enlargement,biagi2020entangling} naturally allow for notably higher photon generation efficiencies assisted by a directionality of the underlying optical phase matching, however, they typically contribute with many spectral and temporal modes simultaneously. A multi-mode spectral nature of a free-space SPDC is typically unavoidable and, in addition, its certification by interferometric measurements is in feasible bandwidth regimes extremely challenging. A conventional enhancement of modeness in these systems corresponds to a severe filtration resulting in a small two-photon coupling efficiencies accompanied by a high susceptibility of observability of QNG to residual noise. Despite significant advancements in demonstrations of operation of narrow-band SPDC sources utilizing optical cavities have been demonstrated~\cite{haase2009,scholz2009,fekete2013,rambach2016,tsai2018,mottola2020efficient,seri2019quantum}, with most recent works approaching a single-mode regime~\cite{moqanaki2019novel}, a provable observation of a single-mode QNG light still remains an open challenge~\cite{lvovsky2020production}.
A rapid development in realization of spontaneous four-wave mixing (SFWM) photon sources with warm atomic vapors over the past decade brought experimentally very feasible demonstrations of a rich variety of spectrally narrow-band and provably nonclassical light with strongly sub-Poissonian and anti-bunched statistics promising a natural applicability for interaction with target atomic ensembles and storage in quantum memories due to their spectral compatibility ~\cite{chen2008,willis2010,ding2012,shu2016subnatural,lee2016,zhu2017bright,podhora2017nonclassical,zugenmaier2018,wang2018,park2018time,mika2020high}. However, to unambiguously surpass the QNG thresholds~\cite{filip2011detecting}, the combination of a residual two-photon noise component and a two-photon coupling efficiency, which are partially competing requirements, have to be established at an unprecedented level~\cite{filip2011detecting,lvovsky2020production}. At the same time, the temporal and spectral modeness of the heralded fields from atomic ensembles with emission spectra dominated by a Doppler broadening in the heralded two-photon regime remains an open question. While the competition between the SFWM gain, absorption losses, and incoherent scattering has been in schemes employing atomic lambda-level configuration widely suppressed with the help of electromagnetically-induced transparency for near resonant anti-Stokes fields, this approach limits feasible spectral bandwidths to several MHz and sets an effective maximal rate of photon generation in the single-mode regime. Another notable approach employs a ladder electronic level scheme, where effects of large Doppler shifts can be conveniently mitigated in a counter-propagating geometry and, in addition, it naturally avoids the dominant noise contribution - Raman noise~\cite{ding2012,lee2016,park2018time}. We believe, it could present a feasible experimental alternative to the scheme presented here.

Here we present a demonstration of heralded emission of QNG light from warm atoms in a single optical mode by the generation of photon pairs in a SFWM in a configuration which avoids any additional frequency filtering of the spectrally matched bi-photons and, simultaneously, allows for suppression of losses of the near-resonant heralding anti-Stokes field. The measured QNG depth up to 43~\% of optical loss opens directions to applications in atomic experiments based on this SFWM source.


\section*{Results}

\begin{figure}[!th]
\begin{center}
\includegraphics[width=1\columnwidth]{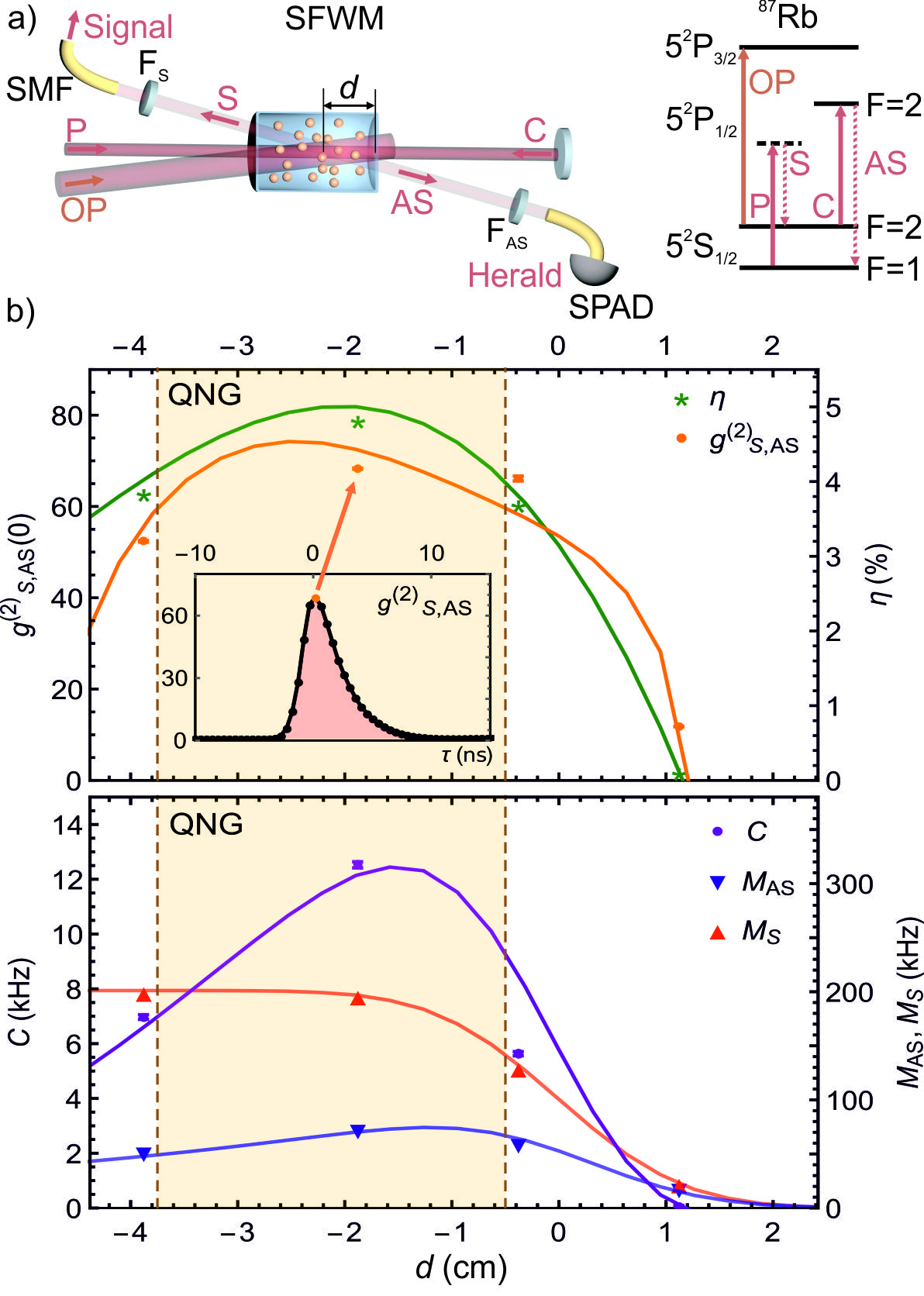}
\caption{SFWM source and basic characteristics of the generated bi-photons. a) shows a simplified excitation geometry realized with a single laser frequency with counter-propagating collimated pump (P) and coupling (C) beams. Distance $d$ is defined between the center of the intersection of excitation and observation spatial modes and the cell output viewport along the pump beam propagation direction. The optical pumping (OP) beam effectively covers this interaction area. The Stokes (S) and anti-Stokes (AS) fields are selected by a combination of Fabry-Pérot cavities, Glan-Thompson polarizers, and optical cut-off filters, marked together as F$_{\rm S}$ and F$_{\rm AS}$, respectively. Spatial modes are set by the coupling into single-mode fibers (SMF) in a two opposite directions. The energy level scheme of $^{87}$Rb depicts the realized SFWM process corresponding to the emission of off-resonant Stokes and of near-resonant anti-Stokes fields  for the excitation laser set close to the $5{\rm S}_{1/2}({\rm F}=2)\leftrightarrow 5{\rm P}_{1/2}({\rm F}=2)$ resonance.
The graphs in b) show basic characteristics of generated biphotons including two-photon coupling efficiency $\eta_{\rm 2ph}$ and observed maxima of the intensity correlation function $g^{(2)}_{\rm S,AS}(0)$ between single photon avalanche diodes (SPAD) in the heralding - anti-Stokes channel and in the signal - Stokes channel. The inset shows an example of $g^{2}_{\rm S, AS}(\tau)$.  Stokes $M_{\rm S}$, anti-Stokes $M_{\rm AS}$, and coincidence $C$ photo-detection rates are shown as red, blue, and magenta data points, respectively. For comparison, QNG marks an area where observation of QNG features is expected. }
\label{fig:block_scheme}
\end{center}
\end{figure}

The feasibility of a single-mode QNG regime based on the conditional detection of photonic state from SFWM in a warm atomic ensemble necessitates suppression of several phenomena contributing with noise and losses to typical realizations of sources of nonclassical light with warm atoms~\cite{podhora2017nonclassical,shu2016subnatural}. It requires tedious simultaneous optimization of SFWM source parameters. These include the matching of the coherence properties and of the frequency bandwidths, optimization of a two-mode nonclassical correlations and two-photon coupling efficiencies, which all have a crucial impact on the feasibility of detection of the challenging QNG states. The presented implementation based on the process of SFWM in a warm $^{87}$Rb vapor can benefit from a unique feasibility of combination of these properties, while its underlaying fundamental scheme is technically simple and corresponds to a double-$\Lambda$ configuration of electronic energy levels with a single counter-propagating laser excitation~\cite{kolchin2006generation,podhora2017nonclassical}. Fig.~\ref{fig:block_scheme} depicts a simplified experimental scheme and main source characteristics. We have implemented several crucial improvements of our scheme~\cite{mika2020high} which contributed to the simultaneous enhancement of the overall two-photon coupling efficiency~$\eta_{\rm 2ph}=P_{\rm C}/{P_{\rm AS}}$ to up to $9.0\pm 0.2$~\% complemented by a more than a tenfold improvement of generated Stokes $M_{\rm S}$ and anti-Stokes $M_{\rm AS}$ photon detection rates and by a twofold increase in photon coincidence rates to $C=7.7\pm 0.1$~kHz evaluated for the coincidence time window $T_{\rm bin}=486$~ps. At the same time, sufficiently high nonclassical correlations between Stokes and anti-Stokes fields on the order of $g^{(2)}_{\rm S,AS}(0)\sim 50$, observable for low excitation powers, could be maintained. The combination of the interaction area corresponding to the spatial overlap of excitation lasers and detection mode within the atomic ensemble in the proximity of the vapor cell viewport~\cite{mika2020high}, efficient auxiliary optical pumping (OP) with a help of atomic polarization preserving coating of vapor cell~\cite{shu2016subnatural}, and particular polarization, spectral, and spatial filtering (F$_{\rm S}$, F$_{\rm AS}$), provide conditions for sufficient two-photon coupling efficiencies and, simultaneously, allow for the feasibility of generation of very high nonclassical correlations and low noise contribution in both Stokes and anti-Stokes fields.

\subsection*{Optimal regime of photon correlations}

Photon characteristics and correlations of light generated in the presented SFWM source, including the two-photon coupling efficiency $\eta_{\rm 2ph}$ and $g^2_{\rm S, AS}(0)$ should be theoretically relatively robust with respect to several conventional excitation parameters including atomic temperature, single-photon detuning of the excitation laser, and its optical power. However, in a regime of a finite loss and excess noise in the Stokes and anti-Stokes fields, the optimization of $g^2_{\rm S, AS}(0)$ tends to favor a low photon rate due to the random nature of background noise and quadratic scaling of its contribution to false coincidences~\cite{mika2018generation}. This is competing with the practical feasibility of high rate of three-photon detection events corresponding to the heralding detection on single-photon avalanche photodiode (SPAD) in the anti-Stokes detector and simultaneous multi-photon detection $P_{2+}$ in the Stokes channel with sufficient statistics. In addition, for a given temperature and atomic optical depth, the position of the interaction area has a paramount impact on the absorption losses of anti-Stokes field and on the corresponding source characteristics. Its precise spatial alignment in the proximity of the output viewport of atomic vapor cell allows for the suppression of losses of the heralding near-resonant anti-Stokes field and simultaneously for its broad spectral width. The frequency spectrum of the two-photon correlations gets purified by the attenuation of the resonant scattering and Raman noise in the Stokes channel, where light scattered within the interaction area traverses through the long ensemble of atoms in the same cell. The observed dependence of the source parameters on the spatial position of the interaction region $d$ shown in the~Fig.~\ref{fig:block_scheme}-b) demonstrates the strong effect of the anti-Stokes filtering and corresponding losses on the normalized intensity correlation function $g^{(2)}_{\rm S, AS}(0)$, two-photon coupling efficiency $\eta$, and on detectable photon rates. The simulations depicted as solid curves include an independently estimated spatial overlap of atoms with Gaussian excitation and observation optical modes, with the scale of the resulting dependencies of photon detection rates fitted to Stokes $M_{\rm S}$ and anti-Stokes $M_{\rm AS}$ data. The paramount role of the spatial alignment $d$ in a large-bandwidth SFWM source is strongly related to the corresponding possibility of preservation of high nonclassical correlations only on very fast timescales of a few nanoseconds, a limit imposed by decoherence rates of collective excitations of thermal atoms~\cite{mitchell2000dynamics}. We note that the observed spatial dependence strongly depends on the interaction length itself, which has been in our case set to a full width at half maximum of $18\pm2$ mm along the excitation direction, well below the value corresponding to a SFWM phase mismatch of $\pi$~\cite{podhora2017nonclassical}. We believe that the proximity of the interaction area to the output cell viewport together with a fast excitation regime are crucial for achievement of QNG properties with SFWM in warm atomic ensembles. Importantly, this approach avoids commonly employed spectral filtering for enhancement of the $g^2_{\rm S, AS}(0)$ and focuses on the suppression of any intrinsic losses and preservation of temporally short nonclassical correlations intrinsic to warm atom systems.

\subsection*{Temporal coherence}

The potential of efficient interference and deterministic coherent interaction of light is given by its modeness, which effectively relates to a degree of purity of the quantum state in all of its degrees of freedom. While the polarization and spatial close-to a single-mode operation is in our demonstration elementarily guaranteed by the employment of a high extinction ratio Glan-Thompson polarizers and single-mode optical fibers, temporal coherence has to be accessed by additional phase-sensitive interferometric measurements. The measured heralded intensity autocorrelation functions could, in principle, consist of an incoherent mixture of several temporal envelopes corresponding to several spectral components. This is particularly important to study in the case of light generated from warm atomic vapors where a large Doppler broadening results in incoherent spectral widths of more than an order of magnitude larger than natural linewidths of employed transitions of alkali atoms. On the other hand, the observable intensity correlations $g^{(2)}(\tau)_{\rm S,AS}$ already show signatures of intensity profiles much below the lifetime of the employed excited states, which can be attributed to the collective nature of the employed SFWM interaction~\cite{park2018temporal,jeong2020temporal,hsu2021generation}.

\begin{figure}[!t]
\begin{center}
\includegraphics[width=1.\columnwidth]{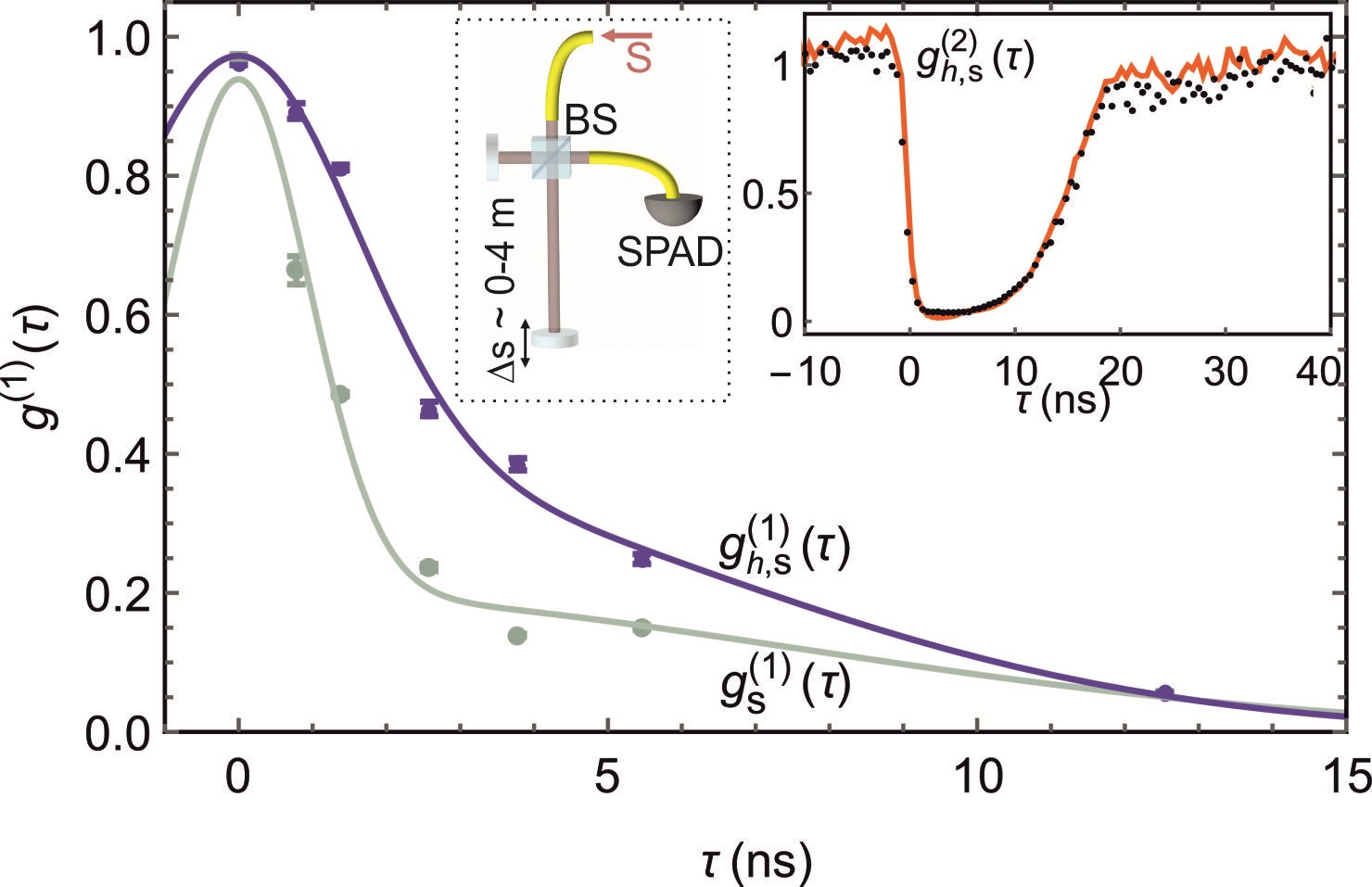}
\caption{Measurements of the first degree of coherence $g^{(1)}_{\rm S}(\tau)$ of Stokes light. The coherence of heralded QNG light $g^{(1)}_{\rm h,S}(\tau)$ and of unconditioned Stokes field is shown as blue and green plots, respectively. The insets show the heralded intensity correlation function $g^{(2)}_{\rm h,S}(\tau)$ evaluated from its direct measurement as black data points and a scheme of Michelson interferometer with a tunable relative path difference $\Delta s$ covering expectable heralded Stokes temporal wave-packets on the order of ten ns. The measured $g^{(2)}_{\rm h,S}(\tau)$ can be plausibly reproduced by simulation employing the measured $g^{(2)}_{\rm S,AS}(\tau)$ and Stokes intensity profile $I_{\rm S}(\tau)$ estimated from the $g^{(2)}_{\rm S}(\tau)$ measurement in an unheralded regime, what suggests a close to unity overlap between the unconditioned and heralded temporal intensity functions of Stokes field.}
\label{fig:modeness}
\end{center}
\end{figure}

We characterize the number of contributing spectral and temporal modes of the generated Stokes-field by analysis of the first degree of coherence $g^{(1)}_{\rm S}(\tau)$. The coherence is measured in a Michelson interferometer with up to approximately 4~m path difference which allows for the relative time delays of more than $\tau\sim 13$~ns corresponding to the extent of observed intensity wavepackets. Fig.~\ref{fig:modeness} shows first-order coherence functions evaluated from the heralded measurement of the QNG Stokes field $g^{(1)}_{\rm h,S}(\tau)$, and $g^{(1)}_{\rm S}(\tau)$ evaluated by including all photons detected in the Stokes channel from the same data acquisition. They can be understood by considering spectral properties of the four-wave-mixing and underlaying phase matching condition, where the third-order susceptibility $\chi^{(3)}$ determining the biphoton spectral waveform includes the enhancement due to the two-photon resonance~\cite{jeong2020temporal,shu2016subnatural}. Crucially, the coherence of the heralded Stokes field is significantly enhanced compared to the non-heralded case, which can be mostly attributed to the residual electromagnetically-induced transparency spectral filtering in the heralding anti-Stokes channel upon passing the atomic ensemble and the corresponding post-selection of Stokes photons through spectral correlations implied by the phase matching conditions~\cite{shu2016subnatural}. The exploitations of analogous spectral correlations have been paramount to the design of SPDC photon sources and their applications~\cite{baek2009spectral}, while they remained largely unexplored in the SFWM domain. A dedicated study of spectral correlation properties in SFWM sources will be part of our next experimental efforts as it reaches beyond the scope of this work. The quantitative comparison of the temporal width of $g^{(1)}_{\rm h, S}(\tau)$ with the corresponding intensity envelope $I_{\rm h, S}(\tau)$ resulted in the temporal mode number $M_{\rm t}=1.43\pm 0.09$. Here $I_{\rm h, S}(\tau)$ has been evaluated indirectly as follows. We used deconvolution of $g^{(2)}_{\rm S}(\tau)$ for accessing $I_{\rm S}(\tau)$ and verified that the following estimation of intensity autocorrelation function of heralded Stokes light $g^{(2)}_{\rm h, S}(\tau)\approx g^{(2)}_{\rm S}/g^{(2)}_{\rm S, AS}$ gives quantitatively good agreement with the measurement. The measured and simulated $g^{(2)}_{\rm h, S}(\tau)$ shown in the inset of the Fig.~\ref{fig:modeness} give high overlap of $96$\,\%, where the residual difference can be partially attributed to the measurement noise. This justifies the consideration of $I_{\rm S,h}(\tau)\sim I_{\rm S}(\tau)$ for the presented evaluation of approximate number of modes in the heralded Stokes field. 
The measured $g^{(1)}_{\rm h, S}(0)=0.963\pm 0.004$ and $g^{(1)}_{\rm S}(0)=0.971\pm 0.003$ differ slightly from ideal value probably due to residual power drifts between the calibration measurements with the coherent laser beam and measurements with Stokes fields.

\subsection*{Quantum non-Gaussian light}

We focus on an unambiguous proof of the QNG character of conditionally generated Stokes field and estimation of its temporal envelope. Here we discuss exclusively the configuration with anti-Stokes trigger and Stokes signal field, because strongly off-resonant Stokes field experiences significantly smaller absorption losses and a corresponding enhancement of a two-photon coupling efficiency. However, we have successfully observed the QNG features also for the inverted configuration with signal field corresponding to a near resonant anti-Stokes emission, see Supplementary information. We also expect feasibility of optimization of the absorption parameters simultaneously for both of the two output channels by the employment of the excitation scheme with independent pump and coupling beams~\cite{shu2016subnatural}.

The QNG of the generated photonic states can be proven by the evaluation of criteria based on the estimated photon number probabilities~\cite{filip2011detecting, lachman2013robustness, straka2014quantum}. In the presented experiment, they are detected using a combination of a fiber-coupled single-photon avalanche photodiodes (SPADs). A single heralding SPAD is employed in the anti-Stokes channel, while a pair of SPADs in the signal-Stokes detection mode is arranged in the single-mode fiber implementation of a Hanbury-Brown-Twiss setup. In the case of the single-photon generation in the presence of noise and losses, the heralded Stokes can be approximately described by
\begin{equation}
\rho \sim P_0 |0\rangle\langle 0| + P_1 |1\rangle\langle 1| + P_{2+} (\sum_{n=2}^{\infty}|n\rangle\langle n|).
\label{eq:generated_state}
\end{equation}
Here $P_0$ and $P_1$ correspond to probabilities of vacuum and single photon states, respectively, and the probability of multiphoton contributions $P_{2+} = 1-P_0-P_1$. Importantly, these probabilities provide a sufficient knowledge of the photon statistics for the estimation of the QNG properties of light emitted by a single-photon source and, at the same time, can be unambiguously evaluated from photon autocorrelation measurements with single-photon detectors~\cite{filip2011detecting, lachman2013robustness, jevzek2011experimental, straka2014quantum}. The sufficient condition for proving the QNG properties of the detected states can be then evaluated as~\cite{lachman2013robustness,straka2014quantum}
\begin{equation}
P_{2+}< \frac{2}{3} P_{1}^3,
\label{eq:QNG_condition}
\end{equation}
which holds for states with $P_{2+}\ll 1$. Note, that this criterion remains reliable also for multi-mode states of light.

\begin{figure}[!t]
\begin{center}
\includegraphics[width=1.\columnwidth]{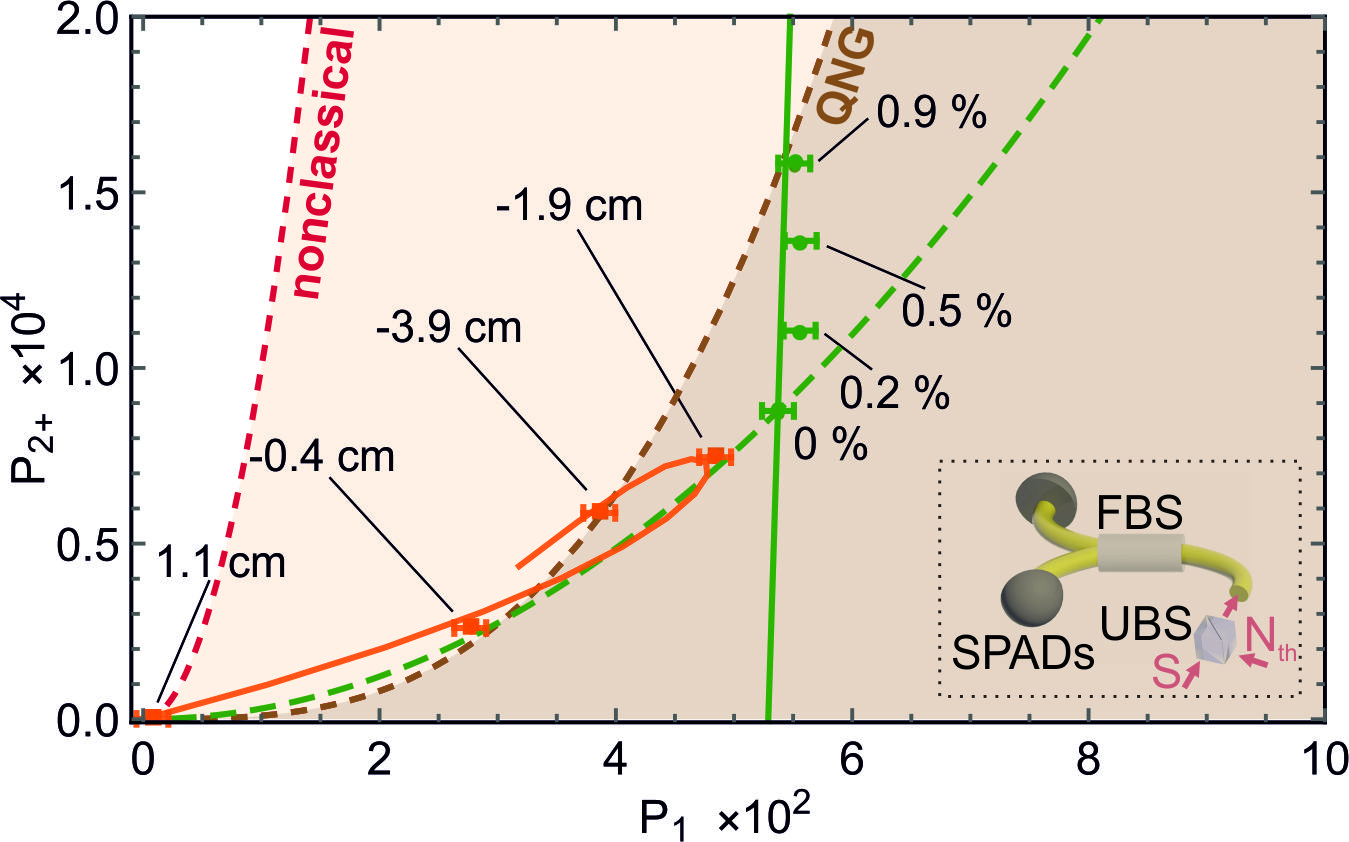}
\caption{Observed QNG features parameterized by photon number probabilities $P_1$ and $P_{2+}$ of a single and multi-photon events observed in the Stokes mode, respectively. The orange data points show raw photodetection probabilities with the corresponding simulation depicted by the orange solid curve. The region of observability of QNG properties corresponds to the shaded areas with the dashed thresholds given by the Eq.~\ref{eq:QNG_condition}. The QNG is unambiguously proved, despite the large sensitivity of their detectability in the presented feasible range of two-photon coupling efficiencies and corresponding Stokes single photon probabilities $P_1$. This sensitivity can be well characterized and quantified  by measurement of a QNG depth, which can be accessed by introduction of an excess thermal noise~\cite{straka2014quantum,lachman2019faithful}. The resulting thermalization trajectory for the interaction position of the interaction area $d=-1.9$~cm is shown as green data points with the thermal noise to Stokes signal ratio ${\rm NSR}=N_{\rm th}/S$ explicitly indicated. The corresponding theoretical model is derived by considering mixing of the generated QNG state with an ideal thermal light. The red dashed curve shows the evaluation of the dependence of the observed QNG features on bare losses in the Stokes mode. The red dashed curve depicts the threshold on the observability of the nonclassical character of measured photon statistics by conventional estimation of the sub-Poissonian value of $\alpha$~parameter. The photon statistics is accessed by detection arrangement consisting of the balanced fiber splitter (FBS) with a fiber-coupled SPAD in each output port shown in the inset. Optional addition of thermal light has been implemented using the unbalanced T:R=90:10\,\% beam-splitter (UBS).}
\label{fig:QNG}
\end{center}
\end{figure}

The evaluations of QNG features for the presented spatial position characterization are depicted in the Fig.~\ref{fig:QNG}. The coincidence time window has been optimized in steps corresponding to an integer multiples of the fundamental resolution of employed time-tagging device of 81~ps to minimize the value $P_{2+}/P_1^3$, with resulting optimal value $T_{\rm bin}=5.67$~ns. Generated photon statistical distributions surpass the criteria~(\ref{eq:QNG_condition}) for an interval of the positions of the interaction region $d$ from $-5$ to $-38$~mm, with a measured peak value at $d=-19\pm 2$~mm corresponding unambiguously to the QNG state. The observed trajectory reflects a clear asymmetry of the observable Stokes and anti-Stokes losses with respect to the position $d$, which can be understood by considering a dominant suppression of the heralding efficiency and of the two-photo coupling efficiency for negative and positive values of $d$, respectively. The evaluations of the corresponding properties of heralded intensity correlations of Stokes field can be found in Supplementary materials. We analyse the sensitivity of observed QNG features to losses and to excess thermal noise in order to allow for its comparison with other sources and to estimate its feasibility for interactions with target atomic ensembles in the presence of realistic noise contributions. The thermal light with a close to ideal thermal statistics is generated by SFWM process in another atomic cell, where the emission is collected without any conditioning on the anti-Stokes detection. This effectively corresponds to tracing of the anti-Stokes field in the two-mode output of the SFWM and results in close to ideal photon bunching and Bose-Einstein photon number probability distribution~\cite{mika2018generation}. The realized thermal source emission is single-mode and has a relatively narrow frequency bandwidth of about 77~MHz set to allow for observation of close to the ideal bunching value $g^{(2)}(\tau=0)=1.95\pm 0.03$ observable with a finite timing jitter of the employed SPADs. The depth of the generated QNG states is accessed by the measurement of photon statistics of light resulting from mixing of the generated QNG Stokes field with the thermal light at the T:R=90:10\,\% unbalanced beam splitter (UBS). The QNG light is sent to a high transmissivity port to limit its loss. Fig.~\ref{fig:QNG} shows the evaluated trajectories for the state generated in the optimal regime of the $d$-parameter for the case of ideal attenuation and for the case of experimentally emulated thermalization with states with mean thermal photon rate of $\bar{N}_{\rm th}$. The thermalization threshold of observability of QNG properties corresponds to the input thermal noise to Stokes signal ratio ${\rm NSR}=P_{\rm th}/P_1=0.94$~\%, where $P_{\rm th}=N_{\rm th} T_{bin}=(4.8\pm0.4)\times10^{-4}$ is the probability of thermal photon detection in a given time window. The simulation of the effect of a bare linear optical loss in the Stokes channel for the same data point is shown as a green dashed line. It suggests expectable preservation of the QNG observability for up to 43.4~\% of optical loss. 
The residual multiphoton contribution $P_{2+}=(8.8\pm0.1)\times10^{-5}$ is partially given by the inevitable thermal noise contributions to the detected Stokes mode coming from the complementary SFWM geometries~\cite{mika2018generation} and by imperfect spatial matching of the coupled Stokes and anti-Stokes modes.
The error bars shown for all data points presented in the Fig.~\ref{fig:QNG} were evaluated from a set of five consecutive measurements and correspond to a single standard deviation of the evaluated parameters. They thus mostly account for a short term statistical uncertainties and don't include long term drifts of experimental parameters on timescales of days or repeatability of the spatial alignment, which were estimated independently to about $\pm0.3~\%$ in the two-photon coupling efficiency $\eta$. This corresponds to the relative fluctuation of the mutual spatial alignment of two fiber-coupled detection channels of 6~\%, which accounts for a visible shift between the set of green data and the corresponding orange data point at $d=-1.9$~cm with no excess thermal noise. We remind that, besides the crucial role of estimation of QNG depth as general and robust comparison of quality of QNG sources with extensions beyond the optical domain~\cite{podhora2021}, it can be also employed for estimation of tolerance of generated QNG properties against intrinsic noise sources and provide a means for an unambiguous optimization of source parameters.

\section*{Discussion}

The QNG properties of light correspond to a paramount resource in majority of prospective applications of quantum technologies~\cite{lvovsky2020production,michael2016new,hu2019quantum,campagne2020quantum,fluhmann2019encoding}. At the same time, the ability to form hybrid quantum systems by controllable interactions provably operating between many QNG states is crucial for reaching their practically useful limits. We have demonstrated a broadly applicable and accessible experimental methodology for generation and analysis of single-mode QNG states in SFWM process with warm atoms. We have shown that photon pairs emitted from SFWM process can provide a conditional generation of statistical properties that prove a new level of experimental control and, importantly, their temporal coherence properties suggest close-to a single-mode operation. The presented scheme with effectively short warm atomic medium provides a high two-photon coupling efficiency without compromising the amplitude of the nonclassical correlations between anti-Stokes and Stokes fields. At the same time, it allows for high photon rates, a necessary condition for feasibility of observation of QNG of photon fields approaching single-photon Fock states in the presence of any residual noise and for practical applicability of the generated QNG light.  Importantly, the natural availability of the natural thermal light sources based on the same process could be employed for a direct experimental evaluation of thermal depth of generated QNG properties, which confirmed an unprecedentedly low thermal noise contribution in the observed photon statistics of the generated Stokes light field. We remind that we have additionally confirmed that QNG regime is achievable also for the anti-Stokes field, although with a quality and depth significantly compromised by lower two-photon coupling efficiency due to residual absorption of near-resonant photons. We foresee significant enhancements of both heralding and two-photon coupling efficiencies by a bare suppression of the passive optical losses in our experiment, with a feasible two-photon coupling efficiencies of about 32~\%. The presented generation of photonic single-mode QNG states capable of a storage and controllable interaction with a bandwidth compatible target atomic systems~\cite{guo2019,reim2011,wolters2017,kaczmarek2018} represents a crucial step for realization of distribution of quantum non-Gaussianity and its local or distributed processing. Together with advancements in the realization of high-efficiency and low-noise optical quantum memories~\cite{hedges2010efficient,hsiao2018highly,vernaz2018highly,wang2019efficient}, the presented source parameters promise potential for an on-demand QNG light source. Demonstrated close-to single-mode operation can provide the missing fundamental ingredient for an efficient nonlinear interactions with QNG light in atomic ensembles~\cite{li2008enhanced,liu2016large,chen2012demonstration}.

\section*{Methods}

In the following description of the SFWM experimental setup we focus only on parts which we found to be paramount for generation and observation of the single-mode QNG states. The SFWM is driven by a continuous-wave single-frequency output of a Ti:Saphire laser blue-detuned about $\Delta=400$\,MHz from the $5{\rm S}_{1/2}({\rm F}=2)\leftrightarrow 5{\rm P}_{1/2}({\rm F}=2)$ transition. The excitation geometry corresponds to a counter-propagating pump-coupling configuration in a convenient retro-reflection setup. Excitation of atoms prepared in the electronic ground state $|g\rangle=5{\rm S}_{1/2}({\rm F}=1)$ results in the excitation-emission cycles $|g\rangle\rightarrow |a_1\rangle\rightarrow 5\mathrm{S}_{1/2}({\rm F}=2)\rightarrow 5\mathrm{P}_{1/2}({\rm F=2})\rightarrow|g\rangle$, where $|a_1\rangle$ corresponds to a virtual energy level red-detuned from the $5\mathrm{P}_{1/2}({\rm F=2})$ by approximately 6.4~GHz. The temperature of the cell was set to 323\,K and excitation beam power was 100~mW for all measurements except where explicitly stated otherwise.
We remind that a small-angle scattering geometry with~$\alpha_{\rm sc}\sim 1.8\pm 0.1$\,$^\circ$ is crucial for the suppression of decoherence of the collective spin wave due to thermal atomic motion on short timescales~\cite{mitchell2000dynamics}. In addition, the Stokes to anti-Stokes correlations are strongly enhanced by the introduction of an auxiliary optical pumping beam set close to the resonance with the 5S$_{1/2}({\rm F} = 2)\leftrightarrow 5{\rm P}_{3/2}$ transition together with a paraffin coating deposited on the inner surfaces of the vapor cell, which substantially suppress the Raman noise in the anti-Stokes mode~\cite{shu2016subnatural,zhu2017bright}. The polarization and spatial degrees of freedom of generated photon pairs are defined by the phase matching condition, which, in the implemented standing-wave excitation geometry and double-$\lambda$ energy level scheme, results in a convenient counter-propagating detection directions and identical linear polarizations~\cite{kolchin2006generation,podhora2017nonclassical}. The polarization and spatial single modeness of Stokes and anti-Stokes fields is guaranteed by the combination of Glan-Thompson polarizer and single-mode optical fibers in each detection channel. The spatial optimization of a coupling to a pair of single-mode optical fibers is optimized to maximize the mutual spatial mode overlap using an auxiliary off-resonant optical beam passing from the Stokes to the anti-Stokes optical fiber, with the resulting overall coupling efficiency matching well an independently estimated optical losses of about 79~\%. The full width of the waist of the observation Gaussian mode at the position of the interaction area, which is defined as an overlap of the observation and excitation optical modes, corresponds to~$95~\pm 5~\mu$m. Fabry-P\'erot filters in both heralding and signal arms have a FWHM frequency bandwidth of $900\pm10$~MHz and are tuned to maximize the transmission of the anti-Stokes and Stokes fields, respectively.

\section*{Acknowledgements}
J.M., L.L. and R.F. acknowledge the support of the Czech Science Foundation under the project GA21-13265X. L. S. is grateful for national
funding from the MEYS under grant agreement No. 731473 and from the QUANTERA ERA-NET cofund in quantum technologies implemented within the European Union’s Horizon 2020 Programme (project PACE-IN, 8C20004). T. L. is grateful for the Palacky University grant IGA-PrF-2021-006.

\bibliographystyle{apsrev4-2}
\bibliography{phase_interference}

\begin{thebibliography}{76}%
\makeatletter
\providecommand \@ifxundefined [1]{%
 \@ifx{#1\undefined}
}%
\providecommand \@ifnum [1]{%
 \ifnum #1\expandafter \@firstoftwo
 \else \expandafter \@secondoftwo
 \fi
}%
\providecommand \@ifx [1]{%
 \ifx #1\expandafter \@firstoftwo
 \else \expandafter \@secondoftwo
 \fi
}%
\providecommand \natexlab [1]{#1}%
\providecommand \enquote  [1]{``#1''}%
\providecommand \bibnamefont  [1]{#1}%
\providecommand \bibfnamefont [1]{#1}%
\providecommand \citenamefont [1]{#1}%
\providecommand \href@noop [0]{\@secondoftwo}%
\providecommand \href [0]{\begingroup \@sanitize@url \@href}%
\providecommand \@href[1]{\@@startlink{#1}\@@href}%
\providecommand \@@href[1]{\endgroup#1\@@endlink}%
\providecommand \@sanitize@url [0]{\catcode `\\12\catcode `\$12\catcode
  `\&12\catcode `\#12\catcode `\^12\catcode `\_12\catcode `\%12\relax}%
\providecommand \@@startlink[1]{}%
\providecommand \@@endlink[0]{}%
\providecommand \url  [0]{\begingroup\@sanitize@url \@url }%
\providecommand \@url [1]{\endgroup\@href {#1}{\urlprefix }}%
\providecommand \urlprefix  [0]{URL }%
\providecommand \Eprint [0]{\href }%
\providecommand \doibase [0]{https://doi.org/}%
\providecommand \selectlanguage [0]{\@gobble}%
\providecommand \bibinfo  [0]{\@secondoftwo}%
\providecommand \bibfield  [0]{\@secondoftwo}%
\providecommand \translation [1]{[#1]}%
\providecommand \BibitemOpen [0]{}%
\providecommand \bibitemStop [0]{}%
\providecommand \bibitemNoStop [0]{.\EOS\space}%
\providecommand \EOS [0]{\spacefactor3000\relax}%
\providecommand \BibitemShut  [1]{\csname bibitem#1\endcsname}%
\let\auto@bib@innerbib\@empty
\bibitem [{\citenamefont {Kurizki}\ \emph {et~al.}(2015)\citenamefont
  {Kurizki}, \citenamefont {Bertet}, \citenamefont {Kubo}, \citenamefont
  {M{\o}lmer}, \citenamefont {Petrosyan}, \citenamefont {Rabl},\ and\
  \citenamefont {Schmiedmayer}}]{kurizki2015quantum}%
  \BibitemOpen
  \bibfield  {author} {\bibinfo {author} {\bibfnamefont {G.}~\bibnamefont
  {Kurizki}}, \bibinfo {author} {\bibfnamefont {P.}~\bibnamefont {Bertet}},
  \bibinfo {author} {\bibfnamefont {Y.}~\bibnamefont {Kubo}}, \bibinfo {author}
  {\bibfnamefont {K.}~\bibnamefont {M{\o}lmer}}, \bibinfo {author}
  {\bibfnamefont {D.}~\bibnamefont {Petrosyan}}, \bibinfo {author}
  {\bibfnamefont {P.}~\bibnamefont {Rabl}},\ and\ \bibinfo {author}
  {\bibfnamefont {J.}~\bibnamefont {Schmiedmayer}},\ }\href
  {https://www.pnas.org/content/112/13/3866} {\bibfield  {journal} {\bibinfo
  {journal} {Proc. Natl. Acad. Sci. U.S.A.}\ }\textbf {\bibinfo {volume}
  {112}},\ \bibinfo {pages} {3866} (\bibinfo {year} {2015})}\BibitemShut
  {NoStop}%
\bibitem [{\citenamefont {Slussarenko}\ and\ \citenamefont
  {Pryde}(2019)}]{slussarenko2019photonic}%
  \BibitemOpen
  \bibfield  {author} {\bibinfo {author} {\bibfnamefont {S.}~\bibnamefont
  {Slussarenko}}\ and\ \bibinfo {author} {\bibfnamefont {G.~J.}\ \bibnamefont
  {Pryde}},\ }\href {https://aip.scitation.org/doi/10.1063/1.5115814}
  {\bibfield  {journal} {\bibinfo  {journal} {Appl. Phys. Rev.}\ }\textbf
  {\bibinfo {volume} {6}},\ \bibinfo {pages} {041303} (\bibinfo {year}
  {2019})}\BibitemShut {NoStop}%
\bibitem [{\citenamefont {Lvovsky}\ \emph {et~al.}(2020)\citenamefont
  {Lvovsky}, \citenamefont {Grangier}, \citenamefont {Ourjoumtsev},
  \citenamefont {Parigi}, \citenamefont {Sasaki},\ and\ \citenamefont
  {Tualle-Brouri}}]{lvovsky2020production}%
  \BibitemOpen
  \bibfield  {author} {\bibinfo {author} {\bibfnamefont {A.~I.}\ \bibnamefont
  {Lvovsky}}, \bibinfo {author} {\bibfnamefont {P.}~\bibnamefont {Grangier}},
  \bibinfo {author} {\bibfnamefont {A.}~\bibnamefont {Ourjoumtsev}}, \bibinfo
  {author} {\bibfnamefont {V.}~\bibnamefont {Parigi}}, \bibinfo {author}
  {\bibfnamefont {M.}~\bibnamefont {Sasaki}},\ and\ \bibinfo {author}
  {\bibfnamefont {R.}~\bibnamefont {Tualle-Brouri}},\ }\href
  {https://arxiv.org/abs/2006.16985} {\bibfield  {journal} {\bibinfo  {journal}
  {arXiv preprint arXiv:2006.16985}\ } (\bibinfo {year} {2020})}\BibitemShut
  {NoStop}%
\bibitem [{\citenamefont {Hudson}(1974)}]{hudson1974wigner}%
  \BibitemOpen
  \bibfield  {author} {\bibinfo {author} {\bibfnamefont {R.~L.}\ \bibnamefont
  {Hudson}},\ }\href {https://doi.org/10.1016/0034-4877(74)90007-X} {\bibfield
  {journal} {\bibinfo  {journal} {Rep. Math. Phys.}\ }\textbf {\bibinfo
  {volume} {6}},\ \bibinfo {pages} {249} (\bibinfo {year} {1974})}\BibitemShut
  {NoStop}%
\bibitem [{\citenamefont {Walschaers}(2021)}]{walschaers2021non}%
  \BibitemOpen
  \bibfield  {author} {\bibinfo {author} {\bibfnamefont {M.}~\bibnamefont
  {Walschaers}},\ }\href
  {https://journals.aps.org/prxquantum/abstract/10.1103/PRXQuantum.2.030204}
  {\bibfield  {journal} {\bibinfo  {journal} {PRX Quantum}\ }\textbf {\bibinfo
  {volume} {2}},\ \bibinfo {pages} {030204} (\bibinfo {year}
  {2021})}\BibitemShut {NoStop}%
\bibitem [{\citenamefont {McCormick}\ \emph {et~al.}(2019)\citenamefont
  {McCormick}, \citenamefont {Keller}, \citenamefont {Burd}, \citenamefont
  {Wineland}, \citenamefont {Wilson},\ and\ \citenamefont
  {Leibfried}}]{mccormick2019quantum}%
  \BibitemOpen
  \bibfield  {author} {\bibinfo {author} {\bibfnamefont {K.~C.}\ \bibnamefont
  {McCormick}}, \bibinfo {author} {\bibfnamefont {J.}~\bibnamefont {Keller}},
  \bibinfo {author} {\bibfnamefont {S.~C.}\ \bibnamefont {Burd}}, \bibinfo
  {author} {\bibfnamefont {D.~J.}\ \bibnamefont {Wineland}}, \bibinfo {author}
  {\bibfnamefont {A.~C.}\ \bibnamefont {Wilson}},\ and\ \bibinfo {author}
  {\bibfnamefont {D.}~\bibnamefont {Leibfried}},\ }\href
  {https://www.nature.com/articles/s41586-019-1421-y} {\bibfield  {journal}
  {\bibinfo  {journal} {Nature}\ }\textbf {\bibinfo {volume} {572}},\ \bibinfo
  {pages} {86} (\bibinfo {year} {2019})}\BibitemShut {NoStop}%
\bibitem [{\citenamefont {Wolf}\ \emph {et~al.}(2019)\citenamefont {Wolf},
  \citenamefont {Shi}, \citenamefont {Heip}, \citenamefont {Gessner},
  \citenamefont {Pezz{\`e}}, \citenamefont {Smerzi}, \citenamefont {Schulte},
  \citenamefont {Hammerer},\ and\ \citenamefont {Schmidt}}]{wolf2019motional}%
  \BibitemOpen
  \bibfield  {author} {\bibinfo {author} {\bibfnamefont {F.}~\bibnamefont
  {Wolf}}, \bibinfo {author} {\bibfnamefont {C.}~\bibnamefont {Shi}}, \bibinfo
  {author} {\bibfnamefont {J.~C.}\ \bibnamefont {Heip}}, \bibinfo {author}
  {\bibfnamefont {M.}~\bibnamefont {Gessner}}, \bibinfo {author} {\bibfnamefont
  {L.}~\bibnamefont {Pezz{\`e}}}, \bibinfo {author} {\bibfnamefont
  {A.}~\bibnamefont {Smerzi}}, \bibinfo {author} {\bibfnamefont
  {M.}~\bibnamefont {Schulte}}, \bibinfo {author} {\bibfnamefont
  {K.}~\bibnamefont {Hammerer}},\ and\ \bibinfo {author} {\bibfnamefont
  {P.~O.}\ \bibnamefont {Schmidt}},\ }\href
  {https://www.nature.com/articles/s41467-019-10576-4} {\bibfield  {journal}
  {\bibinfo  {journal} {Nat. Commun.}\ }\textbf {\bibinfo {volume} {10}},\
  \bibinfo {pages} {1} (\bibinfo {year} {2019})}\BibitemShut {NoStop}%
\bibitem [{\citenamefont {Michael}\ \emph {et~al.}(2016)\citenamefont
  {Michael}, \citenamefont {Silveri}, \citenamefont {Brierley}, \citenamefont
  {Albert}, \citenamefont {Salmilehto}, \citenamefont {Jiang},\ and\
  \citenamefont {Girvin}}]{michael2016new}%
  \BibitemOpen
  \bibfield  {author} {\bibinfo {author} {\bibfnamefont {M.~H.}\ \bibnamefont
  {Michael}}, \bibinfo {author} {\bibfnamefont {M.}~\bibnamefont {Silveri}},
  \bibinfo {author} {\bibfnamefont {R.~T.}\ \bibnamefont {Brierley}}, \bibinfo
  {author} {\bibfnamefont {V.~V.}\ \bibnamefont {Albert}}, \bibinfo {author}
  {\bibfnamefont {J.}~\bibnamefont {Salmilehto}}, \bibinfo {author}
  {\bibfnamefont {L.}~\bibnamefont {Jiang}},\ and\ \bibinfo {author}
  {\bibfnamefont {S.~M.}\ \bibnamefont {Girvin}},\ }\href
  {https://journals.aps.org/prx/abstract/10.1103/PhysRevX.6.031006} {\bibfield
  {journal} {\bibinfo  {journal} {Phys. Rev. X}\ }\textbf {\bibinfo {volume}
  {6}},\ \bibinfo {pages} {031006} (\bibinfo {year} {2016})}\BibitemShut
  {NoStop}%
\bibitem [{\citenamefont {Hu}\ \emph {et~al.}(2019)\citenamefont {Hu},
  \citenamefont {Ma}, \citenamefont {Cai}, \citenamefont {Mu}, \citenamefont
  {Xu}, \citenamefont {Wang}, \citenamefont {Wu}, \citenamefont {Wang},
  \citenamefont {Song}, \citenamefont {Zou}, \citenamefont {Girvin},
  \citenamefont {Duan},\ and\ \citenamefont {Sun}}]{hu2019quantum}%
  \BibitemOpen
  \bibfield  {author} {\bibinfo {author} {\bibfnamefont {L.}~\bibnamefont
  {Hu}}, \bibinfo {author} {\bibfnamefont {Y.}~\bibnamefont {Ma}}, \bibinfo
  {author} {\bibfnamefont {W.}~\bibnamefont {Cai}}, \bibinfo {author}
  {\bibfnamefont {X.}~\bibnamefont {Mu}}, \bibinfo {author} {\bibfnamefont
  {Y.}~\bibnamefont {Xu}}, \bibinfo {author} {\bibfnamefont {W.}~\bibnamefont
  {Wang}}, \bibinfo {author} {\bibfnamefont {Y.}~\bibnamefont {Wu}}, \bibinfo
  {author} {\bibfnamefont {H.}~\bibnamefont {Wang}}, \bibinfo {author}
  {\bibfnamefont {Y.~P.}\ \bibnamefont {Song}}, \bibinfo {author}
  {\bibfnamefont {C.-L.}\ \bibnamefont {Zou}}, \bibinfo {author} {\bibfnamefont
  {S.~M.}\ \bibnamefont {Girvin}}, \bibinfo {author} {\bibfnamefont {L.~M.}\
  \bibnamefont {Duan}},\ and\ \bibinfo {author} {\bibfnamefont
  {L.}~\bibnamefont {Sun}},\ }\href
  {https://www.nature.com/articles/s41567-018-0414-3} {\bibfield  {journal}
  {\bibinfo  {journal} {Nat. Phys.}\ }\textbf {\bibinfo {volume} {15}},\
  \bibinfo {pages} {503} (\bibinfo {year} {2019})}\BibitemShut {NoStop}%
\bibitem [{\citenamefont {Campagne-Ibarcq}\ \emph {et~al.}(2020)\citenamefont
  {Campagne-Ibarcq}, \citenamefont {Eickbusch}, \citenamefont {Touzard},
  \citenamefont {Zalys-Geller}, \citenamefont {Frattini}, \citenamefont
  {Sivak}, \citenamefont {Reinhold}, \citenamefont {Puri}, \citenamefont
  {Shankar}, \citenamefont {Schoelkopf}, \citenamefont {Frunzio}, \citenamefont
  {Mirrahimi},\ and\ \citenamefont {Devoret}}]{campagne2020quantum}%
  \BibitemOpen
  \bibfield  {author} {\bibinfo {author} {\bibfnamefont {P.}~\bibnamefont
  {Campagne-Ibarcq}}, \bibinfo {author} {\bibfnamefont {A.}~\bibnamefont
  {Eickbusch}}, \bibinfo {author} {\bibfnamefont {S.}~\bibnamefont {Touzard}},
  \bibinfo {author} {\bibfnamefont {E.}~\bibnamefont {Zalys-Geller}}, \bibinfo
  {author} {\bibfnamefont {N.~E.}\ \bibnamefont {Frattini}}, \bibinfo {author}
  {\bibfnamefont {V.~V.}\ \bibnamefont {Sivak}}, \bibinfo {author}
  {\bibfnamefont {P.}~\bibnamefont {Reinhold}}, \bibinfo {author}
  {\bibfnamefont {S.}~\bibnamefont {Puri}}, \bibinfo {author} {\bibfnamefont
  {S.}~\bibnamefont {Shankar}}, \bibinfo {author} {\bibfnamefont {R.~J.}\
  \bibnamefont {Schoelkopf}}, \bibinfo {author} {\bibfnamefont
  {L.}~\bibnamefont {Frunzio}}, \bibinfo {author} {\bibfnamefont
  {M.}~\bibnamefont {Mirrahimi}},\ and\ \bibinfo {author} {\bibfnamefont
  {M.~H.}\ \bibnamefont {Devoret}},\ }\href
  {https://www.nature.com/articles/s41586-020-2603-3} {\bibfield  {journal}
  {\bibinfo  {journal} {Nature}\ }\textbf {\bibinfo {volume} {584}},\ \bibinfo
  {pages} {368} (\bibinfo {year} {2020})}\BibitemShut {NoStop}%
\bibitem [{\citenamefont {Fl{\"u}hmann}\ \emph {et~al.}(2019)\citenamefont
  {Fl{\"u}hmann}, \citenamefont {Nguyen}, \citenamefont {Marinelli},
  \citenamefont {Negnevitsky}, \citenamefont {Mehta},\ and\ \citenamefont
  {Home}}]{fluhmann2019encoding}%
  \BibitemOpen
  \bibfield  {author} {\bibinfo {author} {\bibfnamefont {C.}~\bibnamefont
  {Fl{\"u}hmann}}, \bibinfo {author} {\bibfnamefont {T.~L.}\ \bibnamefont
  {Nguyen}}, \bibinfo {author} {\bibfnamefont {M.}~\bibnamefont {Marinelli}},
  \bibinfo {author} {\bibfnamefont {V.}~\bibnamefont {Negnevitsky}}, \bibinfo
  {author} {\bibfnamefont {K.}~\bibnamefont {Mehta}},\ and\ \bibinfo {author}
  {\bibfnamefont {J.~P.}\ \bibnamefont {Home}},\ }\href
  {https://www.nature.com/articles/s41586-019-0960-6} {\bibfield  {journal}
  {\bibinfo  {journal} {Nature}\ }\textbf {\bibinfo {volume} {566}},\ \bibinfo
  {pages} {513} (\bibinfo {year} {2019})}\BibitemShut {NoStop}%
\bibitem [{\citenamefont {Lvovsky}\ \emph {et~al.}(2001)\citenamefont
  {Lvovsky}, \citenamefont {Hansen}, \citenamefont {Aichele}, \citenamefont
  {Benson}, \citenamefont {Mlynek},\ and\ \citenamefont
  {Schiller}}]{lvovsky2001quantum}%
  \BibitemOpen
  \bibfield  {author} {\bibinfo {author} {\bibfnamefont {A.~I.}\ \bibnamefont
  {Lvovsky}}, \bibinfo {author} {\bibfnamefont {H.}~\bibnamefont {Hansen}},
  \bibinfo {author} {\bibfnamefont {T.}~\bibnamefont {Aichele}}, \bibinfo
  {author} {\bibfnamefont {O.}~\bibnamefont {Benson}}, \bibinfo {author}
  {\bibfnamefont {J.}~\bibnamefont {Mlynek}},\ and\ \bibinfo {author}
  {\bibfnamefont {S.}~\bibnamefont {Schiller}},\ }\href
  {https://journals.aps.org/prl/abstract/10.1103/PhysRevLett.87.050402}
  {\bibfield  {journal} {\bibinfo  {journal} {Phys. Rev. Lett.}\ }\textbf
  {\bibinfo {volume} {87}},\ \bibinfo {pages} {050402} (\bibinfo {year}
  {2001})}\BibitemShut {NoStop}%
\bibitem [{\citenamefont {Zavatta}\ \emph {et~al.}(2004)\citenamefont
  {Zavatta}, \citenamefont {Viciani},\ and\ \citenamefont
  {Bellini}}]{zavatta2004tomographic}%
  \BibitemOpen
  \bibfield  {author} {\bibinfo {author} {\bibfnamefont {A.}~\bibnamefont
  {Zavatta}}, \bibinfo {author} {\bibfnamefont {S.}~\bibnamefont {Viciani}},\
  and\ \bibinfo {author} {\bibfnamefont {M.}~\bibnamefont {Bellini}},\ }\href
  {https://arxiv.org/abs/quant-ph/0406090} {\bibfield  {journal} {\bibinfo
  {journal} {Phys. Rev. A}\ }\textbf {\bibinfo {volume} {70}},\ \bibinfo
  {pages} {053821} (\bibinfo {year} {2004})}\BibitemShut {NoStop}%
\bibitem [{\citenamefont {Ourjoumtsev}\ \emph
  {et~al.}(2006{\natexlab{a}})\citenamefont {Ourjoumtsev}, \citenamefont
  {Tualle-Brouri}, \citenamefont {Laurat},\ and\ \citenamefont
  {Grangier}}]{ourjoumtsev2006generating}%
  \BibitemOpen
  \bibfield  {author} {\bibinfo {author} {\bibfnamefont {A.}~\bibnamefont
  {Ourjoumtsev}}, \bibinfo {author} {\bibfnamefont {R.}~\bibnamefont
  {Tualle-Brouri}}, \bibinfo {author} {\bibfnamefont {J.}~\bibnamefont
  {Laurat}},\ and\ \bibinfo {author} {\bibfnamefont {P.}~\bibnamefont
  {Grangier}},\ }\href {https://pubmed.ncbi.nlm.nih.gov/16527930/} {\bibfield
  {journal} {\bibinfo  {journal} {Science}\ }\textbf {\bibinfo {volume}
  {312}},\ \bibinfo {pages} {83} (\bibinfo {year}
  {2006}{\natexlab{a}})}\BibitemShut {NoStop}%
\bibitem [{\citenamefont {Ourjoumtsev}\ \emph
  {et~al.}(2006{\natexlab{b}})\citenamefont {Ourjoumtsev}, \citenamefont
  {Tualle-Brouri},\ and\ \citenamefont {Grangier}}]{ourjoumtsev2006quantum}%
  \BibitemOpen
  \bibfield  {author} {\bibinfo {author} {\bibfnamefont {A.}~\bibnamefont
  {Ourjoumtsev}}, \bibinfo {author} {\bibfnamefont {R.}~\bibnamefont
  {Tualle-Brouri}},\ and\ \bibinfo {author} {\bibfnamefont {P.}~\bibnamefont
  {Grangier}},\ }\href
  {https://journals.aps.org/prl/abstract/10.1103/PhysRevLett.96.213601}
  {\bibfield  {journal} {\bibinfo  {journal} {Phys. Rev. Lett}\ }\textbf
  {\bibinfo {volume} {96}},\ \bibinfo {pages} {213601} (\bibinfo {year}
  {2006}{\natexlab{b}})}\BibitemShut {NoStop}%
\bibitem [{\citenamefont {Miwa}\ \emph {et~al.}(2014)\citenamefont {Miwa},
  \citenamefont {Yoshikawa}, \citenamefont {Iwata}, \citenamefont {Endo},
  \citenamefont {Marek}, \citenamefont {Filip}, \citenamefont {van Loock},\
  and\ \citenamefont {Furusawa}}]{miwa2014exploring}%
  \BibitemOpen
  \bibfield  {author} {\bibinfo {author} {\bibfnamefont {Y.}~\bibnamefont
  {Miwa}}, \bibinfo {author} {\bibfnamefont {J.-i.}\ \bibnamefont {Yoshikawa}},
  \bibinfo {author} {\bibfnamefont {N.}~\bibnamefont {Iwata}}, \bibinfo
  {author} {\bibfnamefont {M.}~\bibnamefont {Endo}}, \bibinfo {author}
  {\bibfnamefont {P.}~\bibnamefont {Marek}}, \bibinfo {author} {\bibfnamefont
  {R.}~\bibnamefont {Filip}}, \bibinfo {author} {\bibfnamefont
  {P.}~\bibnamefont {van Loock}},\ and\ \bibinfo {author} {\bibfnamefont
  {A.}~\bibnamefont {Furusawa}},\ }\href
  {https://journals.aps.org/prl/abstract/10.1103/PhysRevLett.113.013601}
  {\bibfield  {journal} {\bibinfo  {journal} {Phys. Rev. Lett}\ }\textbf
  {\bibinfo {volume} {113}},\ \bibinfo {pages} {013601} (\bibinfo {year}
  {2014})}\BibitemShut {NoStop}%
\bibitem [{\citenamefont {Bimbard}\ \emph {et~al.}(2014)\citenamefont
  {Bimbard}, \citenamefont {Boddeda}, \citenamefont {Vitrant}, \citenamefont
  {Grankin}, \citenamefont {Parigi}, \citenamefont {Stanojevic}, \citenamefont
  {Ourjoumtsev},\ and\ \citenamefont {Grangier}}]{bimbard2014homodyne}%
  \BibitemOpen
  \bibfield  {author} {\bibinfo {author} {\bibfnamefont {E.}~\bibnamefont
  {Bimbard}}, \bibinfo {author} {\bibfnamefont {R.}~\bibnamefont {Boddeda}},
  \bibinfo {author} {\bibfnamefont {N.}~\bibnamefont {Vitrant}}, \bibinfo
  {author} {\bibfnamefont {A.}~\bibnamefont {Grankin}}, \bibinfo {author}
  {\bibfnamefont {V.}~\bibnamefont {Parigi}}, \bibinfo {author} {\bibfnamefont
  {J.}~\bibnamefont {Stanojevic}}, \bibinfo {author} {\bibfnamefont
  {A.}~\bibnamefont {Ourjoumtsev}},\ and\ \bibinfo {author} {\bibfnamefont
  {P.}~\bibnamefont {Grangier}},\ }\href
  {https://journals.aps.org/prl/abstract/10.1103/PhysRevLett.112.033601}
  {\bibfield  {journal} {\bibinfo  {journal} {Phys. Rev. Lett.}\ }\textbf
  {\bibinfo {volume} {112}},\ \bibinfo {pages} {033601} (\bibinfo {year}
  {2014})}\BibitemShut {NoStop}%
\bibitem [{\citenamefont {Le~Jeannic}\ \emph {et~al.}(2018)\citenamefont
  {Le~Jeannic}, \citenamefont {Cavaill{\`e}s}, \citenamefont {Huang},
  \citenamefont {Filip},\ and\ \citenamefont {Laurat}}]{le2018slowing}%
  \BibitemOpen
  \bibfield  {author} {\bibinfo {author} {\bibfnamefont {H.}~\bibnamefont
  {Le~Jeannic}}, \bibinfo {author} {\bibfnamefont {A.}~\bibnamefont
  {Cavaill{\`e}s}}, \bibinfo {author} {\bibfnamefont {K.}~\bibnamefont
  {Huang}}, \bibinfo {author} {\bibfnamefont {R.}~\bibnamefont {Filip}},\ and\
  \bibinfo {author} {\bibfnamefont {J.}~\bibnamefont {Laurat}},\ }\href
  {https://journals.aps.org/prl/abstract/10.1103/PhysRevLett.120.073603}
  {\bibfield  {journal} {\bibinfo  {journal} {Phys. Rev. Lett}\ }\textbf
  {\bibinfo {volume} {120}},\ \bibinfo {pages} {073603} (\bibinfo {year}
  {2018})}\BibitemShut {NoStop}%
\bibitem [{\citenamefont {Zapletal}\ \emph {et~al.}(2021)\citenamefont
  {Zapletal}, \citenamefont {Darras}, \citenamefont {Le~Jeannic}, \citenamefont
  {Cavaill{\`e}s}, \citenamefont {Guccione}, \citenamefont {Laurat},\ and\
  \citenamefont {Filip}}]{zapletal2021experimental}%
  \BibitemOpen
  \bibfield  {author} {\bibinfo {author} {\bibfnamefont {P.}~\bibnamefont
  {Zapletal}}, \bibinfo {author} {\bibfnamefont {T.}~\bibnamefont {Darras}},
  \bibinfo {author} {\bibfnamefont {H.}~\bibnamefont {Le~Jeannic}}, \bibinfo
  {author} {\bibfnamefont {A.}~\bibnamefont {Cavaill{\`e}s}}, \bibinfo {author}
  {\bibfnamefont {G.}~\bibnamefont {Guccione}}, \bibinfo {author}
  {\bibfnamefont {J.}~\bibnamefont {Laurat}},\ and\ \bibinfo {author}
  {\bibfnamefont {R.}~\bibnamefont {Filip}},\ }\href
  {https://www.osapublishing.org/optica/fulltext.cfm?uri=optica-8-5-743&id=451164}
  {\bibfield  {journal} {\bibinfo  {journal} {Optica}\ }\textbf {\bibinfo
  {volume} {8}},\ \bibinfo {pages} {743} (\bibinfo {year} {2021})}\BibitemShut
  {NoStop}%
\bibitem [{\citenamefont {Ra}\ \emph {et~al.}(2020)\citenamefont {Ra},
  \citenamefont {Dufour}, \citenamefont {Walschaers}, \citenamefont {Jacquard},
  \citenamefont {Michel}, \citenamefont {Fabre},\ and\ \citenamefont
  {Treps}}]{ra2020non}%
  \BibitemOpen
  \bibfield  {author} {\bibinfo {author} {\bibfnamefont {Y.-S.}\ \bibnamefont
  {Ra}}, \bibinfo {author} {\bibfnamefont {A.}~\bibnamefont {Dufour}}, \bibinfo
  {author} {\bibfnamefont {M.}~\bibnamefont {Walschaers}}, \bibinfo {author}
  {\bibfnamefont {C.}~\bibnamefont {Jacquard}}, \bibinfo {author}
  {\bibfnamefont {T.}~\bibnamefont {Michel}}, \bibinfo {author} {\bibfnamefont
  {C.}~\bibnamefont {Fabre}},\ and\ \bibinfo {author} {\bibfnamefont
  {N.}~\bibnamefont {Treps}},\ }\href
  {https://www.nature.com/articles/s41567-019-0726-y} {\bibfield  {journal}
  {\bibinfo  {journal} {Nat. Phys.}\ }\textbf {\bibinfo {volume} {16}},\
  \bibinfo {pages} {144} (\bibinfo {year} {2020})}\BibitemShut {NoStop}%
\bibitem [{\citenamefont {Hacker}\ \emph {et~al.}(2019)\citenamefont {Hacker},
  \citenamefont {Welte}, \citenamefont {Daiss}, \citenamefont {Shaukat},
  \citenamefont {Ritter}, \citenamefont {Li},\ and\ \citenamefont
  {Rempe}}]{hacker2019deterministic}%
  \BibitemOpen
  \bibfield  {author} {\bibinfo {author} {\bibfnamefont {B.}~\bibnamefont
  {Hacker}}, \bibinfo {author} {\bibfnamefont {S.}~\bibnamefont {Welte}},
  \bibinfo {author} {\bibfnamefont {S.}~\bibnamefont {Daiss}}, \bibinfo
  {author} {\bibfnamefont {A.}~\bibnamefont {Shaukat}}, \bibinfo {author}
  {\bibfnamefont {S.}~\bibnamefont {Ritter}}, \bibinfo {author} {\bibfnamefont
  {L.}~\bibnamefont {Li}},\ and\ \bibinfo {author} {\bibfnamefont
  {G.}~\bibnamefont {Rempe}},\ }\href
  {https://www.nature.com/articles/s41566-018-0339-5} {\bibfield  {journal}
  {\bibinfo  {journal} {Nat. Photonics}\ }\textbf {\bibinfo {volume} {13}},\
  \bibinfo {pages} {110} (\bibinfo {year} {2019})}\BibitemShut {NoStop}%
\bibitem [{\citenamefont {Schleich}(2011)}]{schleich2011quantum}%
  \BibitemOpen
  \bibfield  {author} {\bibinfo {author} {\bibfnamefont {W.~P.}\ \bibnamefont
  {Schleich}},\ }\href@noop {} {\emph {\bibinfo {title} {Quantum optics in
  phase space}}}\ (\bibinfo  {publisher} {John Wiley \& Sons},\ \bibinfo {year}
  {2011})\BibitemShut {NoStop}%
\bibitem [{\citenamefont {Straka}\ \emph {et~al.}(2014)\citenamefont {Straka},
  \citenamefont {Predojevi{\'c}}, \citenamefont {Huber}, \citenamefont
  {Lachman}, \citenamefont {Butschek}, \citenamefont {Mikov{\'a}},
  \citenamefont {Mi{\v{c}}uda}, \citenamefont {Solomon}, \citenamefont {Weihs},
  \citenamefont {Je{\v{z}}ek} \emph {et~al.}}]{straka2014quantum}%
  \BibitemOpen
  \bibfield  {author} {\bibinfo {author} {\bibfnamefont {I.}~\bibnamefont
  {Straka}}, \bibinfo {author} {\bibfnamefont {A.}~\bibnamefont
  {Predojevi{\'c}}}, \bibinfo {author} {\bibfnamefont {T.}~\bibnamefont
  {Huber}}, \bibinfo {author} {\bibfnamefont {L.}~\bibnamefont {Lachman}},
  \bibinfo {author} {\bibfnamefont {L.}~\bibnamefont {Butschek}}, \bibinfo
  {author} {\bibfnamefont {M.}~\bibnamefont {Mikov{\'a}}}, \bibinfo {author}
  {\bibfnamefont {M.}~\bibnamefont {Mi{\v{c}}uda}}, \bibinfo {author}
  {\bibfnamefont {G.~S.}\ \bibnamefont {Solomon}}, \bibinfo {author}
  {\bibfnamefont {G.}~\bibnamefont {Weihs}}, \bibinfo {author} {\bibfnamefont
  {M.}~\bibnamefont {Je{\v{z}}ek}}, \emph {et~al.},\ }\href
  {https://journals.aps.org/prl/abstract/10.1103/PhysRevLett.113.223603}
  {\bibfield  {journal} {\bibinfo  {journal} {Phys. Rev. Lett.}\ }\textbf
  {\bibinfo {volume} {113}},\ \bibinfo {pages} {223603} (\bibinfo {year}
  {2014})}\BibitemShut {NoStop}%
\bibitem [{\citenamefont {Je{\v{z}}ek}\ \emph {et~al.}(2011)\citenamefont
  {Je{\v{z}}ek}, \citenamefont {Straka}, \citenamefont {Mi{\v{c}}uda},
  \citenamefont {Du{\v{s}}ek}, \citenamefont {Fiur{\'a}{\v{s}}ek},\ and\
  \citenamefont {Filip}}]{jevzek2011experimental}%
  \BibitemOpen
  \bibfield  {author} {\bibinfo {author} {\bibfnamefont {M.}~\bibnamefont
  {Je{\v{z}}ek}}, \bibinfo {author} {\bibfnamefont {I.}~\bibnamefont {Straka}},
  \bibinfo {author} {\bibfnamefont {M.}~\bibnamefont {Mi{\v{c}}uda}}, \bibinfo
  {author} {\bibfnamefont {M.}~\bibnamefont {Du{\v{s}}ek}}, \bibinfo {author}
  {\bibfnamefont {J.}~\bibnamefont {Fiur{\'a}{\v{s}}ek}},\ and\ \bibinfo
  {author} {\bibfnamefont {R.}~\bibnamefont {Filip}},\ }\href
  {https://journals.aps.org/prl/abstract/10.1103/PhysRevLett.107.213602}
  {\bibfield  {journal} {\bibinfo  {journal} {Phys. Rev. Lett.}\ }\textbf
  {\bibinfo {volume} {107}},\ \bibinfo {pages} {213602} (\bibinfo {year}
  {2011})}\BibitemShut {NoStop}%
\bibitem [{\citenamefont {Higginbottom}\ \emph {et~al.}(2016)\citenamefont
  {Higginbottom}, \citenamefont {Slodi{\v{c}}ka}, \citenamefont {Araneda},
  \citenamefont {Lachman}, \citenamefont {Filip}, \citenamefont {Hennrich},\
  and\ \citenamefont {Blatt}}]{higginbottom2016pure}%
  \BibitemOpen
  \bibfield  {author} {\bibinfo {author} {\bibfnamefont {D.~B.}\ \bibnamefont
  {Higginbottom}}, \bibinfo {author} {\bibfnamefont {L.}~\bibnamefont
  {Slodi{\v{c}}ka}}, \bibinfo {author} {\bibfnamefont {G.}~\bibnamefont
  {Araneda}}, \bibinfo {author} {\bibfnamefont {L.}~\bibnamefont {Lachman}},
  \bibinfo {author} {\bibfnamefont {R.}~\bibnamefont {Filip}}, \bibinfo
  {author} {\bibfnamefont {M.}~\bibnamefont {Hennrich}},\ and\ \bibinfo
  {author} {\bibfnamefont {R.}~\bibnamefont {Blatt}},\ }\href
  {https://iopscience.iop.org/article/10.1088/1367-2630/18/9/093038} {\bibfield
   {journal} {\bibinfo  {journal} {New J. Phys.}\ }\textbf {\bibinfo {volume}
  {18}},\ \bibinfo {pages} {093038} (\bibinfo {year} {2016})}\BibitemShut
  {NoStop}%
\bibitem [{\citenamefont {Je{\v{z}}ek}\ \emph {et~al.}(2012)\citenamefont
  {Je{\v{z}}ek}, \citenamefont {Tipsmark}, \citenamefont {Dong}, \citenamefont
  {Fiur{\'a}{\v{s}}ek}, \citenamefont {Mi{\v{s}}ta~Jr}, \citenamefont {Filip},\
  and\ \citenamefont {Andersen}}]{jevzek2012experimental}%
  \BibitemOpen
  \bibfield  {author} {\bibinfo {author} {\bibfnamefont {M.}~\bibnamefont
  {Je{\v{z}}ek}}, \bibinfo {author} {\bibfnamefont {A.}~\bibnamefont
  {Tipsmark}}, \bibinfo {author} {\bibfnamefont {R.}~\bibnamefont {Dong}},
  \bibinfo {author} {\bibfnamefont {J.}~\bibnamefont {Fiur{\'a}{\v{s}}ek}},
  \bibinfo {author} {\bibfnamefont {L.}~\bibnamefont {Mi{\v{s}}ta~Jr}},
  \bibinfo {author} {\bibfnamefont {R.}~\bibnamefont {Filip}},\ and\ \bibinfo
  {author} {\bibfnamefont {U.~L.}\ \bibnamefont {Andersen}},\ }\href
  {https://journals.aps.org/pra/abstract/10.1103/PhysRevA.86.043813} {\bibfield
   {journal} {\bibinfo  {journal} {Phys. Rev. A}\ }\textbf {\bibinfo {volume}
  {86}},\ \bibinfo {pages} {043813} (\bibinfo {year} {2012})}\BibitemShut
  {NoStop}%
\bibitem [{\citenamefont {Lasota}\ \emph {et~al.}(2017)\citenamefont {Lasota},
  \citenamefont {Filip},\ and\ \citenamefont {Usenko}}]{lasota2017sufficiency}%
  \BibitemOpen
  \bibfield  {author} {\bibinfo {author} {\bibfnamefont {M.}~\bibnamefont
  {Lasota}}, \bibinfo {author} {\bibfnamefont {R.}~\bibnamefont {Filip}},\ and\
  \bibinfo {author} {\bibfnamefont {V.~C.}\ \bibnamefont {Usenko}},\ }\href
  {https://journals.aps.org/pra/abstract/10.1103/PhysRevA.96.012301} {\bibfield
   {journal} {\bibinfo  {journal} {Phys. Rev. A}\ }\textbf {\bibinfo {volume}
  {96}},\ \bibinfo {pages} {012301} (\bibinfo {year} {2017})}\BibitemShut
  {NoStop}%
\bibitem [{\citenamefont {Rakhubovsky}\ and\ \citenamefont
  {Filip}(2017)}]{rakhubovsky2017photon}%
  \BibitemOpen
  \bibfield  {author} {\bibinfo {author} {\bibfnamefont {A.~A.}\ \bibnamefont
  {Rakhubovsky}}\ and\ \bibinfo {author} {\bibfnamefont {R.}~\bibnamefont
  {Filip}},\ }\href {https://www.nature.com/articles/srep46764} {\bibfield
  {journal} {\bibinfo  {journal} {Sci. Rep.}\ }\textbf {\bibinfo {volume}
  {7}},\ \bibinfo {pages} {1} (\bibinfo {year} {2017})}\BibitemShut {NoStop}%
\bibitem [{\citenamefont {Niset}\ \emph {et~al.}(2009)\citenamefont {Niset},
  \citenamefont {Fiur{\'a}{\v{s}}ek},\ and\ \citenamefont
  {Cerf}}]{niset2009no}%
  \BibitemOpen
  \bibfield  {author} {\bibinfo {author} {\bibfnamefont {J.}~\bibnamefont
  {Niset}}, \bibinfo {author} {\bibfnamefont {J.}~\bibnamefont
  {Fiur{\'a}{\v{s}}ek}},\ and\ \bibinfo {author} {\bibfnamefont {N.~J.}\
  \bibnamefont {Cerf}},\ }\href
  {https://journals.aps.org/prl/abstract/10.1103/PhysRevLett.102.120501}
  {\bibfield  {journal} {\bibinfo  {journal} {Phys. Rev. Lett.}\ }\textbf
  {\bibinfo {volume} {102}},\ \bibinfo {pages} {120501} (\bibinfo {year}
  {2009})}\BibitemShut {NoStop}%
\bibitem [{\citenamefont {Eisert}\ \emph {et~al.}(2002)\citenamefont {Eisert},
  \citenamefont {Scheel},\ and\ \citenamefont {Plenio}}]{eisert2002distilling}%
  \BibitemOpen
  \bibfield  {author} {\bibinfo {author} {\bibfnamefont {J.}~\bibnamefont
  {Eisert}}, \bibinfo {author} {\bibfnamefont {S.}~\bibnamefont {Scheel}},\
  and\ \bibinfo {author} {\bibfnamefont {M.~B.}\ \bibnamefont {Plenio}},\
  }\href {https://journals.aps.org/prl/abstract/10.1103/PhysRevLett.89.137903}
  {\bibfield  {journal} {\bibinfo  {journal} {Phys. Rev. Lett.}\ }\textbf
  {\bibinfo {volume} {89}},\ \bibinfo {pages} {137903} (\bibinfo {year}
  {2002})}\BibitemShut {NoStop}%
\bibitem [{\citenamefont {Ghose}\ and\ \citenamefont
  {Sanders}(2007)}]{ghose2007non}%
  \BibitemOpen
  \bibfield  {author} {\bibinfo {author} {\bibfnamefont {S.}~\bibnamefont
  {Ghose}}\ and\ \bibinfo {author} {\bibfnamefont {B.~C.}\ \bibnamefont
  {Sanders}},\ }\href {https://arxiv.org/abs/quant-ph/0606026} {\bibfield
  {journal} {\bibinfo  {journal} {J. Mod. Opt.}\ }\textbf {\bibinfo {volume}
  {54}},\ \bibinfo {pages} {855} (\bibinfo {year} {2007})}\BibitemShut
  {NoStop}%
\bibitem [{\citenamefont {Ohliger}\ \emph {et~al.}(2010)\citenamefont
  {Ohliger}, \citenamefont {Kieling},\ and\ \citenamefont
  {Eisert}}]{ohliger2010limitations}%
  \BibitemOpen
  \bibfield  {author} {\bibinfo {author} {\bibfnamefont {M.}~\bibnamefont
  {Ohliger}}, \bibinfo {author} {\bibfnamefont {K.}~\bibnamefont {Kieling}},\
  and\ \bibinfo {author} {\bibfnamefont {J.}~\bibnamefont {Eisert}},\ }\href
  {https://journals.aps.org/pra/abstract/10.1103/PhysRevA.82.042336} {\bibfield
   {journal} {\bibinfo  {journal} {Phys. Rev. A}\ }\textbf {\bibinfo {volume}
  {82}},\ \bibinfo {pages} {042336} (\bibinfo {year} {2010})}\BibitemShut
  {NoStop}%
\bibitem [{\citenamefont {Lachman}\ \emph {et~al.}(2019)\citenamefont
  {Lachman}, \citenamefont {Straka}, \citenamefont {Hlou{\v{s}}ek},
  \citenamefont {Je{\v{z}}ek},\ and\ \citenamefont
  {Filip}}]{lachman2019faithful}%
  \BibitemOpen
  \bibfield  {author} {\bibinfo {author} {\bibfnamefont {L.}~\bibnamefont
  {Lachman}}, \bibinfo {author} {\bibfnamefont {I.}~\bibnamefont {Straka}},
  \bibinfo {author} {\bibfnamefont {J.}~\bibnamefont {Hlou{\v{s}}ek}}, \bibinfo
  {author} {\bibfnamefont {M.}~\bibnamefont {Je{\v{z}}ek}},\ and\ \bibinfo
  {author} {\bibfnamefont {R.}~\bibnamefont {Filip}},\ }\href
  {https://journals.aps.org/prl/abstract/10.1103/PhysRevLett.123.043601}
  {\bibfield  {journal} {\bibinfo  {journal} {Phys. Rev. Lett.}\ }\textbf
  {\bibinfo {volume} {123}},\ \bibinfo {pages} {043601} (\bibinfo {year}
  {2019})}\BibitemShut {NoStop}%
\bibitem [{\citenamefont {Sychev}\ \emph {et~al.}(2017)\citenamefont {Sychev},
  \citenamefont {Ulanov}, \citenamefont {Pushkina}, \citenamefont {Richards},
  \citenamefont {Fedorov},\ and\ \citenamefont
  {Lvovsky}}]{sychev2017enlargement}%
  \BibitemOpen
  \bibfield  {author} {\bibinfo {author} {\bibfnamefont {D.~V.}\ \bibnamefont
  {Sychev}}, \bibinfo {author} {\bibfnamefont {A.~E.}\ \bibnamefont {Ulanov}},
  \bibinfo {author} {\bibfnamefont {A.~A.}\ \bibnamefont {Pushkina}}, \bibinfo
  {author} {\bibfnamefont {M.~W.}\ \bibnamefont {Richards}}, \bibinfo {author}
  {\bibfnamefont {I.~A.}\ \bibnamefont {Fedorov}},\ and\ \bibinfo {author}
  {\bibfnamefont {A.~I.}\ \bibnamefont {Lvovsky}},\ }\href
  {https://www.nature.com/articles/nphoton.2017.57} {\bibfield  {journal}
  {\bibinfo  {journal} {Nat. Photonics}\ }\textbf {\bibinfo {volume} {11}},\
  \bibinfo {pages} {379} (\bibinfo {year} {2017})}\BibitemShut {NoStop}%
\bibitem [{\citenamefont {Biagi}\ \emph {et~al.}(2020)\citenamefont {Biagi},
  \citenamefont {Costanzo}, \citenamefont {Bellini},\ and\ \citenamefont
  {Zavatta}}]{biagi2020entangling}%
  \BibitemOpen
  \bibfield  {author} {\bibinfo {author} {\bibfnamefont {N.}~\bibnamefont
  {Biagi}}, \bibinfo {author} {\bibfnamefont {L.~S.}\ \bibnamefont {Costanzo}},
  \bibinfo {author} {\bibfnamefont {M.}~\bibnamefont {Bellini}},\ and\ \bibinfo
  {author} {\bibfnamefont {A.}~\bibnamefont {Zavatta}},\ }\href
  {https://journals.aps.org/prl/abstract/10.1103/PhysRevLett.124.033604}
  {\bibfield  {journal} {\bibinfo  {journal} {Phys. Rev. Lett.}\ }\textbf
  {\bibinfo {volume} {124}},\ \bibinfo {pages} {033604} (\bibinfo {year}
  {2020})}\BibitemShut {NoStop}%
\bibitem [{\citenamefont {Haase}\ \emph {et~al.}(2009)\citenamefont {Haase},
  \citenamefont {Piro}, \citenamefont {Eschner},\ and\ \citenamefont
  {Mitchell}}]{haase2009}%
  \BibitemOpen
  \bibfield  {author} {\bibinfo {author} {\bibfnamefont {A.}~\bibnamefont
  {Haase}}, \bibinfo {author} {\bibfnamefont {N.}~\bibnamefont {Piro}},
  \bibinfo {author} {\bibfnamefont {J.}~\bibnamefont {Eschner}},\ and\ \bibinfo
  {author} {\bibfnamefont {M.~W.}\ \bibnamefont {Mitchell}},\ }\href
  {https://www.osapublishing.org/ol/abstract.cfm?uri=ol-34-1-55} {\bibfield
  {journal} {\bibinfo  {journal} {Opt. Lett.}\ }\textbf {\bibinfo {volume}
  {34}},\ \bibinfo {pages} {55} (\bibinfo {year} {2009})}\BibitemShut {NoStop}%
\bibitem [{\citenamefont {Scholz}\ \emph {et~al.}(2009)\citenamefont {Scholz},
  \citenamefont {Koch},\ and\ \citenamefont {Benson}}]{scholz2009}%
  \BibitemOpen
  \bibfield  {author} {\bibinfo {author} {\bibfnamefont {M.}~\bibnamefont
  {Scholz}}, \bibinfo {author} {\bibfnamefont {L.}~\bibnamefont {Koch}},\ and\
  \bibinfo {author} {\bibfnamefont {O.}~\bibnamefont {Benson}},\ }\href
  {https://journals.aps.org/prl/abstract/10.1103/PhysRevLett.102.063603}
  {\bibfield  {journal} {\bibinfo  {journal} {Phys. Rev. Lett.}\ }\textbf
  {\bibinfo {volume} {102}},\ \bibinfo {pages} {063603} (\bibinfo {year}
  {2009})}\BibitemShut {NoStop}%
\bibitem [{\citenamefont {Fekete}\ \emph {et~al.}(2013)\citenamefont {Fekete},
  \citenamefont {Riel{\"a}nder}, \citenamefont {Cristiani},\ and\ \citenamefont
  {de~Riedmatten}}]{fekete2013}%
  \BibitemOpen
  \bibfield  {author} {\bibinfo {author} {\bibfnamefont {J.}~\bibnamefont
  {Fekete}}, \bibinfo {author} {\bibfnamefont {D.}~\bibnamefont
  {Riel{\"a}nder}}, \bibinfo {author} {\bibfnamefont {M.}~\bibnamefont
  {Cristiani}},\ and\ \bibinfo {author} {\bibfnamefont {H.}~\bibnamefont
  {de~Riedmatten}},\ }\href
  {https://journals.aps.org/prl/abstract/10.1103/PhysRevLett.110.220502}
  {\bibfield  {journal} {\bibinfo  {journal} {Phys. Rev. Lett.}\ }\textbf
  {\bibinfo {volume} {110}},\ \bibinfo {pages} {220502} (\bibinfo {year}
  {2013})}\BibitemShut {NoStop}%
\bibitem [{\citenamefont {Rambach}\ \emph {et~al.}(2016)\citenamefont
  {Rambach}, \citenamefont {Nikolova}, \citenamefont {Weinhold},\ and\
  \citenamefont {White}}]{rambach2016}%
  \BibitemOpen
  \bibfield  {author} {\bibinfo {author} {\bibfnamefont {M.}~\bibnamefont
  {Rambach}}, \bibinfo {author} {\bibfnamefont {A.}~\bibnamefont {Nikolova}},
  \bibinfo {author} {\bibfnamefont {T.~J.}\ \bibnamefont {Weinhold}},\ and\
  \bibinfo {author} {\bibfnamefont {A.~G.}\ \bibnamefont {White}},\ }\href
  {https://aip.scitation.org/doi/10.1063/1.4966915} {\bibfield  {journal}
  {\bibinfo  {journal} {APL Photonics}\ }\textbf {\bibinfo {volume} {1}},\
  \bibinfo {pages} {096101} (\bibinfo {year} {2016})}\BibitemShut {NoStop}%
\bibitem [{\citenamefont {Tsai}\ and\ \citenamefont {Chen}(2018)}]{tsai2018}%
  \BibitemOpen
  \bibfield  {author} {\bibinfo {author} {\bibfnamefont {P.-J.}\ \bibnamefont
  {Tsai}}\ and\ \bibinfo {author} {\bibfnamefont {Y.-C.}\ \bibnamefont
  {Chen}},\ }\href
  {https://iopscience.iop.org/article/10.1088/2058-9565/aa86e7/meta} {\bibfield
   {journal} {\bibinfo  {journal} {Quantum Sci. Technol.}\ }\textbf {\bibinfo
  {volume} {3}},\ \bibinfo {pages} {034005} (\bibinfo {year}
  {2018})}\BibitemShut {NoStop}%
\bibitem [{\citenamefont {Mottola}\ \emph {et~al.}(2020)\citenamefont
  {Mottola}, \citenamefont {Buser}, \citenamefont {M{\"u}ller}, \citenamefont
  {Kroh}, \citenamefont {Ahlrichs}, \citenamefont {Ramelow}, \citenamefont
  {Benson}, \citenamefont {Treutlein},\ and\ \citenamefont
  {Wolters}}]{mottola2020efficient}%
  \BibitemOpen
  \bibfield  {author} {\bibinfo {author} {\bibfnamefont {R.}~\bibnamefont
  {Mottola}}, \bibinfo {author} {\bibfnamefont {G.}~\bibnamefont {Buser}},
  \bibinfo {author} {\bibfnamefont {C.}~\bibnamefont {M{\"u}ller}}, \bibinfo
  {author} {\bibfnamefont {T.}~\bibnamefont {Kroh}}, \bibinfo {author}
  {\bibfnamefont {A.}~\bibnamefont {Ahlrichs}}, \bibinfo {author}
  {\bibfnamefont {S.}~\bibnamefont {Ramelow}}, \bibinfo {author} {\bibfnamefont
  {O.}~\bibnamefont {Benson}}, \bibinfo {author} {\bibfnamefont
  {P.}~\bibnamefont {Treutlein}},\ and\ \bibinfo {author} {\bibfnamefont
  {J.}~\bibnamefont {Wolters}},\ }\href
  {https://www.osapublishing.org/oe/fulltext.cfm?uri=oe-28-3-3159&id=426151}
  {\bibfield  {journal} {\bibinfo  {journal} {Opt. Express}\ }\textbf {\bibinfo
  {volume} {28}},\ \bibinfo {pages} {3159} (\bibinfo {year}
  {2020})}\BibitemShut {NoStop}%
\bibitem [{\citenamefont {Seri}\ \emph {et~al.}(2019)\citenamefont {Seri},
  \citenamefont {Lago-Rivera}, \citenamefont {Lenhard}, \citenamefont
  {Corrielli}, \citenamefont {Osellame}, \citenamefont {Mazzera},\ and\
  \citenamefont {de~Riedmatten}}]{seri2019quantum}%
  \BibitemOpen
  \bibfield  {author} {\bibinfo {author} {\bibfnamefont {A.}~\bibnamefont
  {Seri}}, \bibinfo {author} {\bibfnamefont {D.}~\bibnamefont {Lago-Rivera}},
  \bibinfo {author} {\bibfnamefont {A.}~\bibnamefont {Lenhard}}, \bibinfo
  {author} {\bibfnamefont {G.}~\bibnamefont {Corrielli}}, \bibinfo {author}
  {\bibfnamefont {R.}~\bibnamefont {Osellame}}, \bibinfo {author}
  {\bibfnamefont {M.}~\bibnamefont {Mazzera}},\ and\ \bibinfo {author}
  {\bibfnamefont {H.}~\bibnamefont {de~Riedmatten}},\ }\href
  {https://www.osapublishing.org/abstract.cfm?uri=EQEC-2019-eaeb_1_1}
  {\bibfield  {journal} {\bibinfo  {journal} {Phys. Rev. Lett.}\ }\textbf
  {\bibinfo {volume} {123}},\ \bibinfo {pages} {080502} (\bibinfo {year}
  {2019})}\BibitemShut {NoStop}%
\bibitem [{\citenamefont {Moqanaki}\ \emph {et~al.}(2019)\citenamefont
  {Moqanaki}, \citenamefont {Massa},\ and\ \citenamefont
  {Walther}}]{moqanaki2019novel}%
  \BibitemOpen
  \bibfield  {author} {\bibinfo {author} {\bibfnamefont {A.}~\bibnamefont
  {Moqanaki}}, \bibinfo {author} {\bibfnamefont {F.}~\bibnamefont {Massa}},\
  and\ \bibinfo {author} {\bibfnamefont {P.}~\bibnamefont {Walther}},\ }\href
  {https://aip.scitation.org/doi/10.1063/1.5095616} {\bibfield  {journal}
  {\bibinfo  {journal} {APL Photonics}\ }\textbf {\bibinfo {volume} {4}},\
  \bibinfo {pages} {090804} (\bibinfo {year} {2019})}\BibitemShut {NoStop}%
\bibitem [{\citenamefont {Chen}\ \emph {et~al.}(2008)\citenamefont {Chen},
  \citenamefont {Shi}, \citenamefont {Feng}, \citenamefont {Zhang},\ and\
  \citenamefont {Guo}}]{chen2008}%
  \BibitemOpen
  \bibfield  {author} {\bibinfo {author} {\bibfnamefont {Q.-F.}\ \bibnamefont
  {Chen}}, \bibinfo {author} {\bibfnamefont {B.-S.}\ \bibnamefont {Shi}},
  \bibinfo {author} {\bibfnamefont {M.}~\bibnamefont {Feng}}, \bibinfo {author}
  {\bibfnamefont {Y.-S.}\ \bibnamefont {Zhang}},\ and\ \bibinfo {author}
  {\bibfnamefont {G.-C.}\ \bibnamefont {Guo}},\ }\href
  {https://www.osapublishing.org/oe/fulltext.cfm?uri=oe-16-26-21708&id=175455}
  {\bibfield  {journal} {\bibinfo  {journal} {Opt. Express}\ }\textbf {\bibinfo
  {volume} {16}},\ \bibinfo {pages} {21708} (\bibinfo {year}
  {2008})}\BibitemShut {NoStop}%
\bibitem [{\citenamefont {Willis}\ \emph {et~al.}(2010)\citenamefont {Willis},
  \citenamefont {Becerra}, \citenamefont {Orozco},\ and\ \citenamefont
  {Rolston}}]{willis2010}%
  \BibitemOpen
  \bibfield  {author} {\bibinfo {author} {\bibfnamefont {R.}~\bibnamefont
  {Willis}}, \bibinfo {author} {\bibfnamefont {F.}~\bibnamefont {Becerra}},
  \bibinfo {author} {\bibfnamefont {L.}~\bibnamefont {Orozco}},\ and\ \bibinfo
  {author} {\bibfnamefont {S.}~\bibnamefont {Rolston}},\ }\href
  {https://journals.aps.org/pra/abstract/10.1103/PhysRevA.82.053842} {\bibfield
   {journal} {\bibinfo  {journal} {Phys. Rev. A}\ }\textbf {\bibinfo {volume}
  {82}},\ \bibinfo {pages} {053842} (\bibinfo {year} {2010})}\BibitemShut
  {NoStop}%
\bibitem [{\citenamefont {Ding}\ \emph {et~al.}(2012)\citenamefont {Ding},
  \citenamefont {Zhou}, \citenamefont {Shi}, \citenamefont {Zou},\ and\
  \citenamefont {Guo}}]{ding2012}%
  \BibitemOpen
  \bibfield  {author} {\bibinfo {author} {\bibfnamefont {D.-S.}\ \bibnamefont
  {Ding}}, \bibinfo {author} {\bibfnamefont {Z.-Y.}\ \bibnamefont {Zhou}},
  \bibinfo {author} {\bibfnamefont {B.-S.}\ \bibnamefont {Shi}}, \bibinfo
  {author} {\bibfnamefont {X.-B.}\ \bibnamefont {Zou}},\ and\ \bibinfo {author}
  {\bibfnamefont {G.-C.}\ \bibnamefont {Guo}},\ }\href
  {https://www.osapublishing.org/oe/fulltext.cfm?uri=oe-20-10-11433&id=233067}
  {\bibfield  {journal} {\bibinfo  {journal} {Opt. Express}\ }\textbf {\bibinfo
  {volume} {20}},\ \bibinfo {pages} {11433} (\bibinfo {year}
  {2012})}\BibitemShut {NoStop}%
\bibitem [{\citenamefont {Shu}\ \emph {et~al.}(2016)\citenamefont {Shu},
  \citenamefont {Chen}, \citenamefont {Chow}, \citenamefont {Zhu},
  \citenamefont {Xiao}, \citenamefont {Loy},\ and\ \citenamefont
  {Du}}]{shu2016subnatural}%
  \BibitemOpen
  \bibfield  {author} {\bibinfo {author} {\bibfnamefont {C.}~\bibnamefont
  {Shu}}, \bibinfo {author} {\bibfnamefont {P.}~\bibnamefont {Chen}}, \bibinfo
  {author} {\bibfnamefont {T.~K.~A.}\ \bibnamefont {Chow}}, \bibinfo {author}
  {\bibfnamefont {L.}~\bibnamefont {Zhu}}, \bibinfo {author} {\bibfnamefont
  {Y.}~\bibnamefont {Xiao}}, \bibinfo {author} {\bibfnamefont {M.~M.~T.}\
  \bibnamefont {Loy}},\ and\ \bibinfo {author} {\bibfnamefont {S.}~\bibnamefont
  {Du}},\ }\href {https://www.nature.com/articles/ncomms12783} {\bibfield
  {journal} {\bibinfo  {journal} {Nat. Commun}\ }\textbf {\bibinfo {volume}
  {7}},\ \bibinfo {pages} {1} (\bibinfo {year} {2016})}\BibitemShut {NoStop}%
\bibitem [{\citenamefont {Lee}\ \emph {et~al.}(2016)\citenamefont {Lee},
  \citenamefont {Lee}, \citenamefont {Kim},\ and\ \citenamefont
  {Moon}}]{lee2016}%
  \BibitemOpen
  \bibfield  {author} {\bibinfo {author} {\bibfnamefont {Y.-S.}\ \bibnamefont
  {Lee}}, \bibinfo {author} {\bibfnamefont {S.~M.}\ \bibnamefont {Lee}},
  \bibinfo {author} {\bibfnamefont {H.}~\bibnamefont {Kim}},\ and\ \bibinfo
  {author} {\bibfnamefont {H.~S.}\ \bibnamefont {Moon}},\ }\href
  {https://www.osapublishing.org/oe/fulltext.cfm?uri=oe-24-24-28083&id=355566}
  {\bibfield  {journal} {\bibinfo  {journal} {Opt. Express}\ }\textbf {\bibinfo
  {volume} {24}},\ \bibinfo {pages} {28083} (\bibinfo {year}
  {2016})}\BibitemShut {NoStop}%
\bibitem [{\citenamefont {Zhu}\ \emph {et~al.}(2017)\citenamefont {Zhu},
  \citenamefont {Guo}, \citenamefont {Shu}, \citenamefont {Jeong},\ and\
  \citenamefont {Du}}]{zhu2017bright}%
  \BibitemOpen
  \bibfield  {author} {\bibinfo {author} {\bibfnamefont {L.}~\bibnamefont
  {Zhu}}, \bibinfo {author} {\bibfnamefont {X.}~\bibnamefont {Guo}}, \bibinfo
  {author} {\bibfnamefont {C.}~\bibnamefont {Shu}}, \bibinfo {author}
  {\bibfnamefont {H.}~\bibnamefont {Jeong}},\ and\ \bibinfo {author}
  {\bibfnamefont {S.}~\bibnamefont {Du}},\ }\href
  {https://aip.scitation.org/doi/10.1063/1.4980073} {\bibfield  {journal}
  {\bibinfo  {journal} {Appl. Phys. Lett.}\ }\textbf {\bibinfo {volume}
  {110}},\ \bibinfo {pages} {161101} (\bibinfo {year} {2017})}\BibitemShut
  {NoStop}%
\bibitem [{\citenamefont {Podhora}\ \emph {et~al.}(2017)\citenamefont
  {Podhora}, \citenamefont {Ob{\v{s}}il}, \citenamefont {Straka}, \citenamefont
  {Je{\v{z}}ek},\ and\ \citenamefont
  {Slodi{\v{c}}ka}}]{podhora2017nonclassical}%
  \BibitemOpen
  \bibfield  {author} {\bibinfo {author} {\bibfnamefont {L.}~\bibnamefont
  {Podhora}}, \bibinfo {author} {\bibfnamefont {P.}~\bibnamefont
  {Ob{\v{s}}il}}, \bibinfo {author} {\bibfnamefont {I.}~\bibnamefont {Straka}},
  \bibinfo {author} {\bibfnamefont {M.}~\bibnamefont {Je{\v{z}}ek}},\ and\
  \bibinfo {author} {\bibfnamefont {L.}~\bibnamefont {Slodi{\v{c}}ka}},\ }\href
  {https://www.osapublishing.org/oe/fulltext.cfm?uri=oe-25-25-31230&id=379142}
  {\bibfield  {journal} {\bibinfo  {journal} {Opt. Express}\ }\textbf {\bibinfo
  {volume} {25}},\ \bibinfo {pages} {31230} (\bibinfo {year}
  {2017})}\BibitemShut {NoStop}%
\bibitem [{\citenamefont {Zugenmaier}\ \emph {et~al.}(2018)\citenamefont
  {Zugenmaier}, \citenamefont {Dideriksen}, \citenamefont {S{\o}rensen},
  \citenamefont {Albrecht},\ and\ \citenamefont {Polzik}}]{zugenmaier2018}%
  \BibitemOpen
  \bibfield  {author} {\bibinfo {author} {\bibfnamefont {M.}~\bibnamefont
  {Zugenmaier}}, \bibinfo {author} {\bibfnamefont {K.~B.}\ \bibnamefont
  {Dideriksen}}, \bibinfo {author} {\bibfnamefont {A.~S.}\ \bibnamefont
  {S{\o}rensen}}, \bibinfo {author} {\bibfnamefont {B.}~\bibnamefont
  {Albrecht}},\ and\ \bibinfo {author} {\bibfnamefont {E.~S.}\ \bibnamefont
  {Polzik}},\ }\href {https://www.nature.com/articles/s42005-018-0080-x}
  {\bibfield  {journal} {\bibinfo  {journal} {Commun. Phys}\ }\textbf {\bibinfo
  {volume} {1}},\ \bibinfo {pages} {76} (\bibinfo {year} {2018})}\BibitemShut
  {NoStop}%
\bibitem [{\citenamefont {Wang}\ \emph {et~al.}(2018)\citenamefont {Wang},
  \citenamefont {Gu}, \citenamefont {Yu}, \citenamefont {Wei}, \citenamefont
  {Zhang}, \citenamefont {Gao},\ and\ \citenamefont {Li}}]{wang2018}%
  \BibitemOpen
  \bibfield  {author} {\bibinfo {author} {\bibfnamefont {C.}~\bibnamefont
  {Wang}}, \bibinfo {author} {\bibfnamefont {Y.}~\bibnamefont {Gu}}, \bibinfo
  {author} {\bibfnamefont {Y.}~\bibnamefont {Yu}}, \bibinfo {author}
  {\bibfnamefont {D.}~\bibnamefont {Wei}}, \bibinfo {author} {\bibfnamefont
  {P.}~\bibnamefont {Zhang}}, \bibinfo {author} {\bibfnamefont
  {H.}~\bibnamefont {Gao}},\ and\ \bibinfo {author} {\bibfnamefont
  {F.}~\bibnamefont {Li}},\ }\href
  {https://www.osapublishing.org/col/abstract.cfm?uri=col-16-8-082701}
  {\bibfield  {journal} {\bibinfo  {journal} {Chin. Opt. Lett.}\ }\textbf
  {\bibinfo {volume} {16}},\ \bibinfo {pages} {082701} (\bibinfo {year}
  {2018})}\BibitemShut {NoStop}%
\bibitem [{\citenamefont {Park}\ \emph
  {et~al.}(2018{\natexlab{a}})\citenamefont {Park}, \citenamefont {Jeong},
  \citenamefont {Kim},\ and\ \citenamefont {Moon}}]{park2018time}%
  \BibitemOpen
  \bibfield  {author} {\bibinfo {author} {\bibfnamefont {J.}~\bibnamefont
  {Park}}, \bibinfo {author} {\bibfnamefont {T.}~\bibnamefont {Jeong}},
  \bibinfo {author} {\bibfnamefont {H.}~\bibnamefont {Kim}},\ and\ \bibinfo
  {author} {\bibfnamefont {H.~S.}\ \bibnamefont {Moon}},\ }\href
  {https://journals.aps.org/prl/abstract/10.1103/PhysRevLett.121.263601}
  {\bibfield  {journal} {\bibinfo  {journal} {Phys. Rev. Lett.}\ }\textbf
  {\bibinfo {volume} {121}},\ \bibinfo {pages} {263601} (\bibinfo {year}
  {2018}{\natexlab{a}})}\BibitemShut {NoStop}%
\bibitem [{\citenamefont {Mika}\ and\ \citenamefont
  {Slodi{\v{c}}ka}(2020)}]{mika2020high}%
  \BibitemOpen
  \bibfield  {author} {\bibinfo {author} {\bibfnamefont {J.}~\bibnamefont
  {Mika}}\ and\ \bibinfo {author} {\bibfnamefont {L.}~\bibnamefont
  {Slodi{\v{c}}ka}},\ }\href
  {https://iopscience.iop.org/article/10.1088/1361-6455/ab8717} {\bibfield
  {journal} {\bibinfo  {journal} {J. Phys. B}\ }\textbf {\bibinfo {volume}
  {53}},\ \bibinfo {pages} {145501} (\bibinfo {year} {2020})}\BibitemShut
  {NoStop}%
\bibitem [{\citenamefont {Filip}\ and\ \citenamefont
  {Mi{\v{s}}ta~Jr}(2011)}]{filip2011detecting}%
  \BibitemOpen
  \bibfield  {author} {\bibinfo {author} {\bibfnamefont {R.}~\bibnamefont
  {Filip}}\ and\ \bibinfo {author} {\bibfnamefont {L.}~\bibnamefont
  {Mi{\v{s}}ta~Jr}},\ }\href
  {https://journals.aps.org/prl/abstract/10.1103/PhysRevLett.106.200401}
  {\bibfield  {journal} {\bibinfo  {journal} {Phys. Rev. Lett.}\ }\textbf
  {\bibinfo {volume} {106}},\ \bibinfo {pages} {200401} (\bibinfo {year}
  {2011})}\BibitemShut {NoStop}%
\bibitem [{\citenamefont {Kolchin}\ \emph {et~al.}(2006)\citenamefont
  {Kolchin}, \citenamefont {Du}, \citenamefont {Belthangady}, \citenamefont
  {Yin},\ and\ \citenamefont {Harris}}]{kolchin2006generation}%
  \BibitemOpen
  \bibfield  {author} {\bibinfo {author} {\bibfnamefont {P.}~\bibnamefont
  {Kolchin}}, \bibinfo {author} {\bibfnamefont {S.}~\bibnamefont {Du}},
  \bibinfo {author} {\bibfnamefont {C.}~\bibnamefont {Belthangady}}, \bibinfo
  {author} {\bibfnamefont {G.~Y.}\ \bibnamefont {Yin}},\ and\ \bibinfo {author}
  {\bibfnamefont {S.~E.}\ \bibnamefont {Harris}},\ }\href
  {https://journals.aps.org/prl/abstract/10.1103/PhysRevLett.97.113602}
  {\bibfield  {journal} {\bibinfo  {journal} {Phys. Rev. Lett.}\ }\textbf
  {\bibinfo {volume} {97}},\ \bibinfo {pages} {113602} (\bibinfo {year}
  {2006})}\BibitemShut {NoStop}%
\bibitem [{\citenamefont {Mika}\ \emph {et~al.}(2018)\citenamefont {Mika},
  \citenamefont {Podhora}, \citenamefont {Lachman}, \citenamefont
  {Ob{\v{s}}il}, \citenamefont {Hlou{\v{s}}ek}, \citenamefont {Je{\v{z}}ek},
  \citenamefont {Filip},\ and\ \citenamefont
  {Slodi{\v{c}}ka}}]{mika2018generation}%
  \BibitemOpen
  \bibfield  {author} {\bibinfo {author} {\bibfnamefont {J.}~\bibnamefont
  {Mika}}, \bibinfo {author} {\bibfnamefont {L.}~\bibnamefont {Podhora}},
  \bibinfo {author} {\bibfnamefont {L.}~\bibnamefont {Lachman}}, \bibinfo
  {author} {\bibfnamefont {P.}~\bibnamefont {Ob{\v{s}}il}}, \bibinfo {author}
  {\bibfnamefont {J.}~\bibnamefont {Hlou{\v{s}}ek}}, \bibinfo {author}
  {\bibfnamefont {M.}~\bibnamefont {Je{\v{z}}ek}}, \bibinfo {author}
  {\bibfnamefont {R.}~\bibnamefont {Filip}},\ and\ \bibinfo {author}
  {\bibfnamefont {L.}~\bibnamefont {Slodi{\v{c}}ka}},\ }\href
  {https://iopscience.iop.org/article/10.1088/1367-2630/aadc9d} {\bibfield
  {journal} {\bibinfo  {journal} {New J. Phys.}\ }\textbf {\bibinfo {volume}
  {20}},\ \bibinfo {pages} {093002} (\bibinfo {year} {2018})}\BibitemShut
  {NoStop}%
\bibitem [{\citenamefont {Mitchell}\ \emph {et~al.}(2000)\citenamefont
  {Mitchell}, \citenamefont {Hancox},\ and\ \citenamefont
  {Chiao}}]{mitchell2000dynamics}%
  \BibitemOpen
  \bibfield  {author} {\bibinfo {author} {\bibfnamefont {M.~W.}\ \bibnamefont
  {Mitchell}}, \bibinfo {author} {\bibfnamefont {C.~I.}\ \bibnamefont
  {Hancox}},\ and\ \bibinfo {author} {\bibfnamefont {R.~Y.}\ \bibnamefont
  {Chiao}},\ }\href
  {https://journals.aps.org/pra/abstract/10.1103/PhysRevA.62.043819} {\bibfield
   {journal} {\bibinfo  {journal} {Phys. Rev. A}\ }\textbf {\bibinfo {volume}
  {62}},\ \bibinfo {pages} {043819} (\bibinfo {year} {2000})}\BibitemShut
  {NoStop}%
\bibitem [{\citenamefont {Park}\ \emph
  {et~al.}(2018{\natexlab{b}})\citenamefont {Park}, \citenamefont {Jeong},\
  and\ \citenamefont {Moon}}]{park2018temporal}%
  \BibitemOpen
  \bibfield  {author} {\bibinfo {author} {\bibfnamefont {J.}~\bibnamefont
  {Park}}, \bibinfo {author} {\bibfnamefont {T.}~\bibnamefont {Jeong}},\ and\
  \bibinfo {author} {\bibfnamefont {H.~S.}\ \bibnamefont {Moon}},\ }\href
  {https://www.nature.com/articles/s41598-018-29340-7} {\bibfield  {journal}
  {\bibinfo  {journal} {Sci. Rep.}\ }\textbf {\bibinfo {volume} {8}},\ \bibinfo
  {pages} {1} (\bibinfo {year} {2018}{\natexlab{b}})}\BibitemShut {NoStop}%
\bibitem [{\citenamefont {Jeong}\ and\ \citenamefont
  {Moon}(2020)}]{jeong2020temporal}%
  \BibitemOpen
  \bibfield  {author} {\bibinfo {author} {\bibfnamefont {T.}~\bibnamefont
  {Jeong}}\ and\ \bibinfo {author} {\bibfnamefont {H.~S.}\ \bibnamefont
  {Moon}},\ }\href
  {https://www.osapublishing.org/oe/fulltext.cfm?uri=oe-28-3-3985&id=426406}
  {\bibfield  {journal} {\bibinfo  {journal} {Opt. Express}\ }\textbf {\bibinfo
  {volume} {28}},\ \bibinfo {pages} {3985} (\bibinfo {year}
  {2020})}\BibitemShut {NoStop}%
\bibitem [{\citenamefont {Hsu}\ \emph {et~al.}(2021)\citenamefont {Hsu},
  \citenamefont {Wang}, \citenamefont {Chen}, \citenamefont {Huang},
  \citenamefont {Ke}, \citenamefont {Huang}, \citenamefont {Hung},
  \citenamefont {Chao}, \citenamefont {Hsiao}, \citenamefont {Chen} \emph
  {et~al.}}]{hsu2021generation}%
  \BibitemOpen
  \bibfield  {author} {\bibinfo {author} {\bibfnamefont {C.-Y.}\ \bibnamefont
  {Hsu}}, \bibinfo {author} {\bibfnamefont {Y.-S.}\ \bibnamefont {Wang}},
  \bibinfo {author} {\bibfnamefont {J.-M.}\ \bibnamefont {Chen}}, \bibinfo
  {author} {\bibfnamefont {F.-C.}\ \bibnamefont {Huang}}, \bibinfo {author}
  {\bibfnamefont {Y.-T.}\ \bibnamefont {Ke}}, \bibinfo {author} {\bibfnamefont
  {E.~K.}\ \bibnamefont {Huang}}, \bibinfo {author} {\bibfnamefont
  {W.}~\bibnamefont {Hung}}, \bibinfo {author} {\bibfnamefont {K.-L.}\
  \bibnamefont {Chao}}, \bibinfo {author} {\bibfnamefont {S.-S.}\ \bibnamefont
  {Hsiao}}, \bibinfo {author} {\bibfnamefont {Y.-H.}\ \bibnamefont {Chen}},
  \emph {et~al.},\ }\href
  {https://www.osapublishing.org/oe/fulltext.cfm?uri=oe-29-3-4632&id=447030}
  {\bibfield  {journal} {\bibinfo  {journal} {Opt. Express}\ }\textbf {\bibinfo
  {volume} {29}},\ \bibinfo {pages} {4632} (\bibinfo {year}
  {2021})}\BibitemShut {NoStop}%
\bibitem [{\citenamefont {Baek}\ and\ \citenamefont
  {Kim}(2009)}]{baek2009spectral}%
  \BibitemOpen
  \bibfield  {author} {\bibinfo {author} {\bibfnamefont {S.-Y.}\ \bibnamefont
  {Baek}}\ and\ \bibinfo {author} {\bibfnamefont {Y.-H.}\ \bibnamefont {Kim}},\
  }\href {https://www.osapublishing.org/abstract.cfm?uri=CLEO-2009-JTuD96}
  {\bibfield  {journal} {\bibinfo  {journal} {Phys. Rev. A}\ }\textbf {\bibinfo
  {volume} {80}},\ \bibinfo {pages} {033814} (\bibinfo {year}
  {2009})}\BibitemShut {NoStop}%
\bibitem [{\citenamefont {Lachman}\ and\ \citenamefont
  {Filip}(2013)}]{lachman2013robustness}%
  \BibitemOpen
  \bibfield  {author} {\bibinfo {author} {\bibfnamefont {L.}~\bibnamefont
  {Lachman}}\ and\ \bibinfo {author} {\bibfnamefont {R.}~\bibnamefont
  {Filip}},\ }\href
  {https://journals.aps.org/pra/abstract/10.1103/PhysRevA.88.063841} {\bibfield
   {journal} {\bibinfo  {journal} {Phys. Rev. A}\ }\textbf {\bibinfo {volume}
  {88}},\ \bibinfo {pages} {063841} (\bibinfo {year} {2013})}\BibitemShut
  {NoStop}%
\bibitem [{\citenamefont {Podhora}\ \emph {et~al.}(2021)\citenamefont
  {Podhora}, \citenamefont {Lachman}, \citenamefont {Pham}, \citenamefont
  {Le{\v{s}}und{\'a}k}, \citenamefont {{\v{C}}ip}, \citenamefont
  {Slodi{\v{c}}ka},\ and\ \citenamefont {Filip}}]{podhora2021}%
  \BibitemOpen
  \bibfield  {author} {\bibinfo {author} {\bibfnamefont {L.}~\bibnamefont
  {Podhora}}, \bibinfo {author} {\bibfnamefont {L.}~\bibnamefont {Lachman}},
  \bibinfo {author} {\bibfnamefont {T.}~\bibnamefont {Pham}}, \bibinfo {author}
  {\bibfnamefont {A.}~\bibnamefont {Le{\v{s}}und{\'a}k}}, \bibinfo {author}
  {\bibfnamefont {O.}~\bibnamefont {{\v{C}}ip}}, \bibinfo {author}
  {\bibfnamefont {L.}~\bibnamefont {Slodi{\v{c}}ka}},\ and\ \bibinfo {author}
  {\bibfnamefont {R.}~\bibnamefont {Filip}},\ }\href
  {https://arxiv.org/abs/2111.10129} {\bibfield  {journal} {\bibinfo  {journal}
  {arXiv preprint arXiv:2111.10129}\ } (\bibinfo {year} {2021})}\BibitemShut
  {NoStop}%
\bibitem [{\citenamefont {Guo}\ \emph {et~al.}(2019)\citenamefont {Guo},
  \citenamefont {Feng}, \citenamefont {Yang}, \citenamefont {Yu}, \citenamefont
  {Chen}, \citenamefont {Yuan},\ and\ \citenamefont {Zhang}}]{guo2019}%
  \BibitemOpen
  \bibfield  {author} {\bibinfo {author} {\bibfnamefont {J.}~\bibnamefont
  {Guo}}, \bibinfo {author} {\bibfnamefont {X.}~\bibnamefont {Feng}}, \bibinfo
  {author} {\bibfnamefont {P.}~\bibnamefont {Yang}}, \bibinfo {author}
  {\bibfnamefont {Z.}~\bibnamefont {Yu}}, \bibinfo {author} {\bibfnamefont
  {L.~Q.}\ \bibnamefont {Chen}}, \bibinfo {author} {\bibfnamefont {C.-H.}\
  \bibnamefont {Yuan}},\ and\ \bibinfo {author} {\bibfnamefont
  {W.}~\bibnamefont {Zhang}},\ }\href
  {https://www.nature.com/articles/s41467-018-08118-5} {\bibfield  {journal}
  {\bibinfo  {journal} {Nat. Commun}\ }\textbf {\bibinfo {volume} {10}},\
  \bibinfo {pages} {148} (\bibinfo {year} {2019})}\BibitemShut {NoStop}%
\bibitem [{\citenamefont {Reim}\ \emph {et~al.}(2011)\citenamefont {Reim},
  \citenamefont {Michelberger}, \citenamefont {Lee}, \citenamefont {Nunn},
  \citenamefont {Langford},\ and\ \citenamefont {Walmsley}}]{reim2011}%
  \BibitemOpen
  \bibfield  {author} {\bibinfo {author} {\bibfnamefont {K.~F.}\ \bibnamefont
  {Reim}}, \bibinfo {author} {\bibfnamefont {P.}~\bibnamefont {Michelberger}},
  \bibinfo {author} {\bibfnamefont {K.~C.}\ \bibnamefont {Lee}}, \bibinfo
  {author} {\bibfnamefont {J.}~\bibnamefont {Nunn}}, \bibinfo {author}
  {\bibfnamefont {N.~K.}\ \bibnamefont {Langford}},\ and\ \bibinfo {author}
  {\bibfnamefont {I.~A.}\ \bibnamefont {Walmsley}},\ }\href
  {https://journals.aps.org/prl/abstract/10.1103/PhysRevLett.107.053603}
  {\bibfield  {journal} {\bibinfo  {journal} {Phys. Rev. Lett.}\ }\textbf
  {\bibinfo {volume} {107}},\ \bibinfo {pages} {053603} (\bibinfo {year}
  {2011})}\BibitemShut {NoStop}%
\bibitem [{\citenamefont {Wolters}\ \emph {et~al.}(2017)\citenamefont
  {Wolters}, \citenamefont {Buser}, \citenamefont {Horsley}, \citenamefont
  {B{\'e}guin}, \citenamefont {J{\"o}ckel}, \citenamefont {Jahn}, \citenamefont
  {Warburton},\ and\ \citenamefont {Treutlein}}]{wolters2017}%
  \BibitemOpen
  \bibfield  {author} {\bibinfo {author} {\bibfnamefont {J.}~\bibnamefont
  {Wolters}}, \bibinfo {author} {\bibfnamefont {G.}~\bibnamefont {Buser}},
  \bibinfo {author} {\bibfnamefont {A.}~\bibnamefont {Horsley}}, \bibinfo
  {author} {\bibfnamefont {L.}~\bibnamefont {B{\'e}guin}}, \bibinfo {author}
  {\bibfnamefont {A.}~\bibnamefont {J{\"o}ckel}}, \bibinfo {author}
  {\bibfnamefont {J.-P.}\ \bibnamefont {Jahn}}, \bibinfo {author}
  {\bibfnamefont {R.~J.}\ \bibnamefont {Warburton}},\ and\ \bibinfo {author}
  {\bibfnamefont {P.}~\bibnamefont {Treutlein}},\ }\href
  {https://journals.aps.org/prl/abstract/10.1103/PhysRevLett.119.060502}
  {\bibfield  {journal} {\bibinfo  {journal} {Phys. Rev. Lett.}\ }\textbf
  {\bibinfo {volume} {119}},\ \bibinfo {pages} {060502} (\bibinfo {year}
  {2017})}\BibitemShut {NoStop}%
\bibitem [{\citenamefont {Kaczmarek}\ \emph {et~al.}(2018)\citenamefont
  {Kaczmarek}, \citenamefont {Ledingham}, \citenamefont {Brecht}, \citenamefont
  {Thomas}, \citenamefont {Thekkadath}, \citenamefont {Lazo-Arjona},
  \citenamefont {Munns}, \citenamefont {Poem}, \citenamefont {Feizpour},
  \citenamefont {Saunders} \emph {et~al.}}]{kaczmarek2018}%
  \BibitemOpen
  \bibfield  {author} {\bibinfo {author} {\bibfnamefont {K.~T.}\ \bibnamefont
  {Kaczmarek}}, \bibinfo {author} {\bibfnamefont {P.~M.}\ \bibnamefont
  {Ledingham}}, \bibinfo {author} {\bibfnamefont {B.}~\bibnamefont {Brecht}},
  \bibinfo {author} {\bibfnamefont {S.~E.}\ \bibnamefont {Thomas}}, \bibinfo
  {author} {\bibfnamefont {G.~S.}\ \bibnamefont {Thekkadath}}, \bibinfo
  {author} {\bibfnamefont {O.}~\bibnamefont {Lazo-Arjona}}, \bibinfo {author}
  {\bibfnamefont {J.~H.~D.}\ \bibnamefont {Munns}}, \bibinfo {author}
  {\bibfnamefont {E.}~\bibnamefont {Poem}}, \bibinfo {author} {\bibfnamefont
  {A.}~\bibnamefont {Feizpour}}, \bibinfo {author} {\bibfnamefont {D.~J.}\
  \bibnamefont {Saunders}}, \emph {et~al.},\ }\href
  {https://journals.aps.org/pra/abstract/10.1103/PhysRevA.97.042316} {\bibfield
   {journal} {\bibinfo  {journal} {Phys. Rev. A}\ }\textbf {\bibinfo {volume}
  {97}},\ \bibinfo {pages} {042316} (\bibinfo {year} {2018})}\BibitemShut
  {NoStop}%
\bibitem [{\citenamefont {Hedges}\ \emph {et~al.}(2010)\citenamefont {Hedges},
  \citenamefont {Longdell}, \citenamefont {Li},\ and\ \citenamefont
  {Sellars}}]{hedges2010efficient}%
  \BibitemOpen
  \bibfield  {author} {\bibinfo {author} {\bibfnamefont {M.~P.}\ \bibnamefont
  {Hedges}}, \bibinfo {author} {\bibfnamefont {J.~J.}\ \bibnamefont
  {Longdell}}, \bibinfo {author} {\bibfnamefont {Y.}~\bibnamefont {Li}},\ and\
  \bibinfo {author} {\bibfnamefont {M.~J.}\ \bibnamefont {Sellars}},\ }\href
  {https://www.nature.com/articles/nature09081} {\bibfield  {journal} {\bibinfo
   {journal} {Nature}\ }\textbf {\bibinfo {volume} {465}},\ \bibinfo {pages}
  {1052} (\bibinfo {year} {2010})}\BibitemShut {NoStop}%
\bibitem [{\citenamefont {Hsiao}\ \emph {et~al.}(2018)\citenamefont {Hsiao},
  \citenamefont {Tsai}, \citenamefont {Chen}, \citenamefont {Lin},
  \citenamefont {Hung}, \citenamefont {Lee}, \citenamefont {Chen},
  \citenamefont {Chen}, \citenamefont {Ite},\ and\ \citenamefont
  {Chen}}]{hsiao2018highly}%
  \BibitemOpen
  \bibfield  {author} {\bibinfo {author} {\bibfnamefont {Y.-F.}\ \bibnamefont
  {Hsiao}}, \bibinfo {author} {\bibfnamefont {P.-J.}\ \bibnamefont {Tsai}},
  \bibinfo {author} {\bibfnamefont {H.-S.}\ \bibnamefont {Chen}}, \bibinfo
  {author} {\bibfnamefont {S.-X.}\ \bibnamefont {Lin}}, \bibinfo {author}
  {\bibfnamefont {C.-C.}\ \bibnamefont {Hung}}, \bibinfo {author}
  {\bibfnamefont {C.-H.}\ \bibnamefont {Lee}}, \bibinfo {author} {\bibfnamefont
  {Y.-H.}\ \bibnamefont {Chen}}, \bibinfo {author} {\bibfnamefont {Y.-F.}\
  \bibnamefont {Chen}}, \bibinfo {author} {\bibfnamefont {A.~Y.}\ \bibnamefont
  {Ite}},\ and\ \bibinfo {author} {\bibfnamefont {Y.-C.}\ \bibnamefont
  {Chen}},\ }\href
  {https://journals.aps.org/prl/abstract/10.1103/PhysRevLett.120.183602}
  {\bibfield  {journal} {\bibinfo  {journal} {Phys. Rev. Lett}\ }\textbf
  {\bibinfo {volume} {120}},\ \bibinfo {pages} {183602} (\bibinfo {year}
  {2018})}\BibitemShut {NoStop}%
\bibitem [{\citenamefont {Vernaz-Gris}\ \emph {et~al.}(2018)\citenamefont
  {Vernaz-Gris}, \citenamefont {Huang}, \citenamefont {Cao}, \citenamefont
  {Sheremet},\ and\ \citenamefont {Laurat}}]{vernaz2018highly}%
  \BibitemOpen
  \bibfield  {author} {\bibinfo {author} {\bibfnamefont {P.}~\bibnamefont
  {Vernaz-Gris}}, \bibinfo {author} {\bibfnamefont {K.}~\bibnamefont {Huang}},
  \bibinfo {author} {\bibfnamefont {M.}~\bibnamefont {Cao}}, \bibinfo {author}
  {\bibfnamefont {A.~S.}\ \bibnamefont {Sheremet}},\ and\ \bibinfo {author}
  {\bibfnamefont {J.}~\bibnamefont {Laurat}},\ }\href
  {https://www.nature.com/articles/s41467-017-02775-8} {\bibfield  {journal}
  {\bibinfo  {journal} {Nat. Commun.}\ }\textbf {\bibinfo {volume} {9}},\
  \bibinfo {pages} {1} (\bibinfo {year} {2018})}\BibitemShut {NoStop}%
\bibitem [{\citenamefont {Wang}\ \emph {et~al.}(2019)\citenamefont {Wang},
  \citenamefont {Li}, \citenamefont {Zhang}, \citenamefont {Su}, \citenamefont
  {Zhou}, \citenamefont {Liao}, \citenamefont {Du}, \citenamefont {Yan},\ and\
  \citenamefont {Zhu}}]{wang2019efficient}%
  \BibitemOpen
  \bibfield  {author} {\bibinfo {author} {\bibfnamefont {Y.}~\bibnamefont
  {Wang}}, \bibinfo {author} {\bibfnamefont {J.}~\bibnamefont {Li}}, \bibinfo
  {author} {\bibfnamefont {S.}~\bibnamefont {Zhang}}, \bibinfo {author}
  {\bibfnamefont {K.}~\bibnamefont {Su}}, \bibinfo {author} {\bibfnamefont
  {Y.}~\bibnamefont {Zhou}}, \bibinfo {author} {\bibfnamefont {K.}~\bibnamefont
  {Liao}}, \bibinfo {author} {\bibfnamefont {S.}~\bibnamefont {Du}}, \bibinfo
  {author} {\bibfnamefont {H.}~\bibnamefont {Yan}},\ and\ \bibinfo {author}
  {\bibfnamefont {S.-L.}\ \bibnamefont {Zhu}},\ }\href
  {https://www.nature.com/articles/s41566-019-0368-8} {\bibfield  {journal}
  {\bibinfo  {journal} {Nat. Photonics}\ }\textbf {\bibinfo {volume} {13}},\
  \bibinfo {pages} {346} (\bibinfo {year} {2019})}\BibitemShut {NoStop}%
\bibitem [{\citenamefont {Li}\ \emph {et~al.}(2008)\citenamefont {Li},
  \citenamefont {Yang}, \citenamefont {Cao}, \citenamefont {Zhang},
  \citenamefont {Xie},\ and\ \citenamefont {Wang}}]{li2008enhanced}%
  \BibitemOpen
  \bibfield  {author} {\bibinfo {author} {\bibfnamefont {S.}~\bibnamefont
  {Li}}, \bibinfo {author} {\bibfnamefont {X.}~\bibnamefont {Yang}}, \bibinfo
  {author} {\bibfnamefont {X.}~\bibnamefont {Cao}}, \bibinfo {author}
  {\bibfnamefont {C.}~\bibnamefont {Zhang}}, \bibinfo {author} {\bibfnamefont
  {C.}~\bibnamefont {Xie}},\ and\ \bibinfo {author} {\bibfnamefont
  {H.}~\bibnamefont {Wang}},\ }\href
  {https://journals.aps.org/prl/abstract/10.1103/PhysRevLett.101.073602}
  {\bibfield  {journal} {\bibinfo  {journal} {Phys. Rev. Lett}\ }\textbf
  {\bibinfo {volume} {101}},\ \bibinfo {pages} {073602} (\bibinfo {year}
  {2008})}\BibitemShut {NoStop}%
\bibitem [{\citenamefont {Liu}\ \emph {et~al.}(2016)\citenamefont {Liu},
  \citenamefont {Chen}, \citenamefont {Chen}, \citenamefont {Lo}, \citenamefont
  {Tsai}, \citenamefont {Ite}, \citenamefont {Chen},\ and\ \citenamefont
  {Chen}}]{liu2016large}%
  \BibitemOpen
  \bibfield  {author} {\bibinfo {author} {\bibfnamefont {Z.-Y.}\ \bibnamefont
  {Liu}}, \bibinfo {author} {\bibfnamefont {Y.-H.}\ \bibnamefont {Chen}},
  \bibinfo {author} {\bibfnamefont {Y.-C.}\ \bibnamefont {Chen}}, \bibinfo
  {author} {\bibfnamefont {H.-Y.}\ \bibnamefont {Lo}}, \bibinfo {author}
  {\bibfnamefont {P.-J.}\ \bibnamefont {Tsai}}, \bibinfo {author}
  {\bibfnamefont {A.~Y.}\ \bibnamefont {Ite}}, \bibinfo {author} {\bibfnamefont
  {Y.-C.}\ \bibnamefont {Chen}},\ and\ \bibinfo {author} {\bibfnamefont
  {Y.-F.}\ \bibnamefont {Chen}},\ }\href
  {https://journals.aps.org/prl/abstract/10.1103/PhysRevLett.117.203601}
  {\bibfield  {journal} {\bibinfo  {journal} {Phys. Rev. Lett}\ }\textbf
  {\bibinfo {volume} {117}},\ \bibinfo {pages} {203601} (\bibinfo {year}
  {2016})}\BibitemShut {NoStop}%
\bibitem [{\citenamefont {Chen}\ \emph {et~al.}(2012)\citenamefont {Chen},
  \citenamefont {Lee}, \citenamefont {Hung}, \citenamefont {Chen},
  \citenamefont {Chen},\ and\ \citenamefont {Ite}}]{chen2012demonstration}%
  \BibitemOpen
  \bibfield  {author} {\bibinfo {author} {\bibfnamefont {Y.-H.}\ \bibnamefont
  {Chen}}, \bibinfo {author} {\bibfnamefont {M.-J.}\ \bibnamefont {Lee}},
  \bibinfo {author} {\bibfnamefont {W.}~\bibnamefont {Hung}}, \bibinfo {author}
  {\bibfnamefont {Y.-C.}\ \bibnamefont {Chen}}, \bibinfo {author}
  {\bibfnamefont {Y.-F.}\ \bibnamefont {Chen}},\ and\ \bibinfo {author}
  {\bibfnamefont {A.~Y.}\ \bibnamefont {Ite}},\ }\href
  {https://journals.aps.org/prl/abstract/10.1103/PhysRevLett.108.173603}
  {\bibfield  {journal} {\bibinfo  {journal} {Phys. Rev. Lett}\ }\textbf
  {\bibinfo {volume} {108}},\ \bibinfo {pages} {173603} (\bibinfo {year}
  {2012})}\BibitemShut {NoStop}%
\bibitem [{\citenamefont {Lachman}\ \emph {et~al.}(2016)\citenamefont
  {Lachman}, \citenamefont {Slodi{\v{c}}ka},\ and\ \citenamefont
  {Filip}}]{lachman2016nonclassical}%
  \BibitemOpen
  \bibfield  {author} {\bibinfo {author} {\bibfnamefont {L.}~\bibnamefont
  {Lachman}}, \bibinfo {author} {\bibfnamefont {L.}~\bibnamefont
  {Slodi{\v{c}}ka}},\ and\ \bibinfo {author} {\bibfnamefont {R.}~\bibnamefont
  {Filip}},\ }\href {https://www.nature.com/articles/srep19760} {\bibfield
  {journal} {\bibinfo  {journal} {Sci. Rep.}\ }\textbf {\bibinfo {volume}
  {6}},\ \bibinfo {pages} {1} (\bibinfo {year} {2016})}\BibitemShut {NoStop}%
\end{thebibliography}%

\clearpage

\clearpage

\newpage

\renewcommand{\theequation}{S\arabic{equation}}
\renewcommand{\thefigure}{S\arabic{figure}}
\renewcommand{\thesection}{S\arabic{section}}
\renewcommand{\theHequation}{Supplement.\theequation}
\renewcommand{\theHfigure}{Supplement.\thefigure}
\renewcommand{\bibnumfmt}[1]{[S#1]}
\renewcommand{\citenumfont}[1]{S#1}

\setcounter{equation}{0}
\setcounter{figure}{0}
\setcounter{table}{0}
\setcounter{section}{0}
\setcounter{page}{1} \makeatletter

\section*{Supplementary information: Single-mode Quantum Non-Gaussian Light from Warm Atoms}

\section*{Source optimization}
\label{SI:experimental_implementation}

\begin{figure}[!t]
\begin{center}
\includegraphics[width=1.\columnwidth]{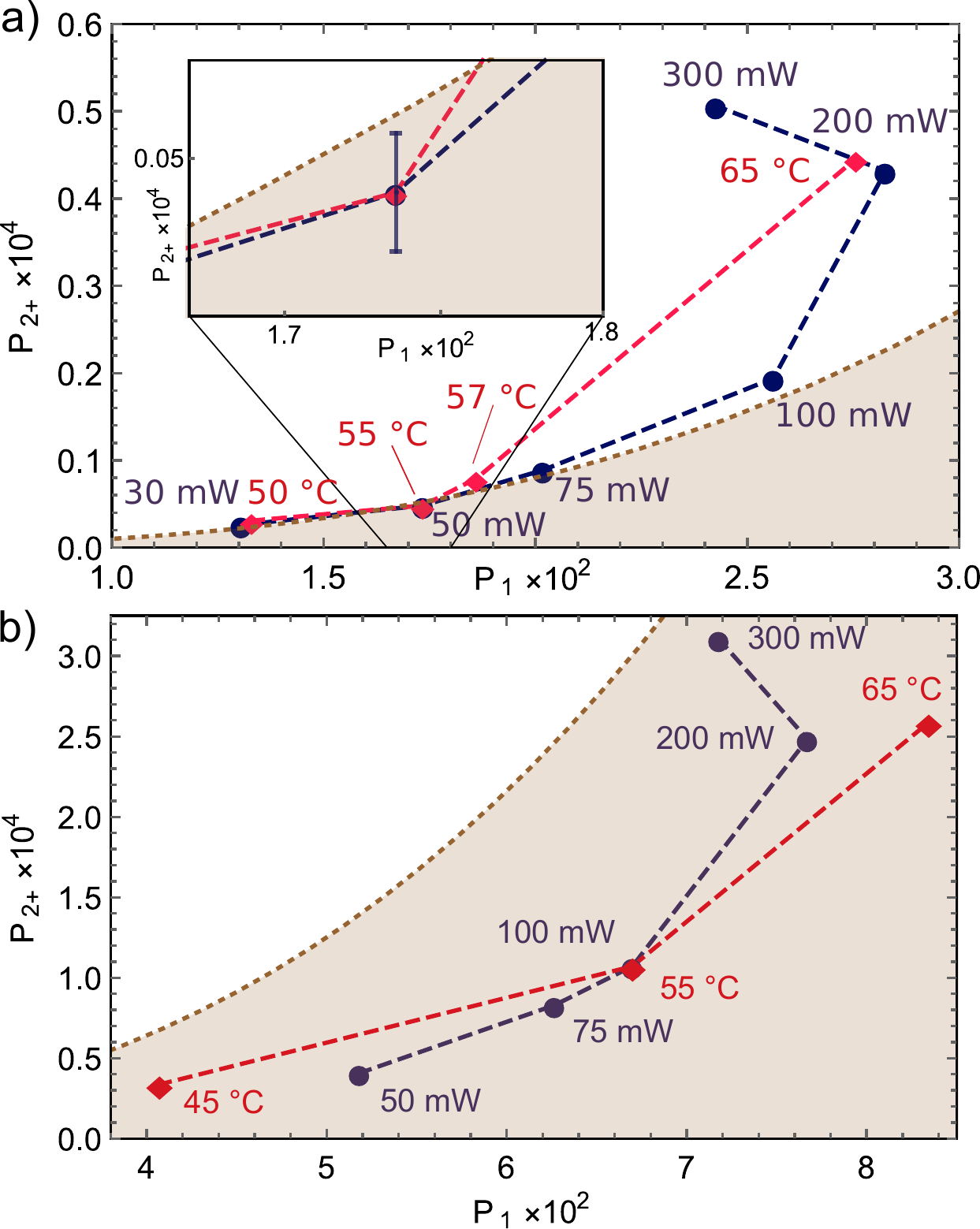}
\caption{The measured dependence of the photon statistics of the heralded fields on the SFWM excitation beam power and temperature of the atomic cell shown with blue circles and red squares, respectively. The dark areas mark the QNG regions. The a) shows the corresponding behaviour for the heralded Stokes field and b) in the heralded anti-Stokes field. In the power dependence measurements, the temperature has been set to $T=55^\circ$\,C. For the temperature dependence measurements, the power has been set to 100~mW and 50~mW for the graphs a) and b), respectively. The dashed lines serve to guide the eye with arrows marking the direction of increase of the studied parameter. The error bars correspond to a single standard deviation evaluated statistically from a set of five measurements. The inset in a) shows the detail of the area close to the QNG threshold around the data point at 50~mW.}
\label{fig:powerTempS}
\end{center}
\end{figure}

The measured power and temperature dependence of the observability of QNG properties from the SFWM source in the heralding anti-Stokes and Stokes regimes are shown in the Fig.~\ref{fig:powerTempS}-a) and b), respectively. The robust dependence of the unambiguous observability of QNG on both the excitation laser power and the number of atoms in the cell for considered anti-Stokes trigger provides a good perspective towards scaling of the generated QNG states to higher genuine QNG features~\cite{lachman2019faithful} and realization of efficient interferometric interactions with target photonic or atomic systems. The presented photon number probabilities were evaluated from measured time-tagged photo-counts with time bin widths set to $T_{\rm bin}=5.67$~ns. The basic scaling of the photon number probabilities $P_1$ and $P_{2+}$ on the number of contributing atoms corresponding to the temperature of the cell $T$ is in both cases dominantly determined by a quadratic behaviour of the random contribution of noise to $P_{\rm 2}$ and linear increase of heralded single-photon signal corresponding to the desired SFWM output~\cite{mika2018generation,shu2016subnatural}. In addition, a nontrivial character manifested by the decrease of the $P_1$ for very high excitation beam powers corresponds to saturation of the enhancement of directionality of collectively emitted photons due to the enhancement of coupling Rabi frequency on timescales of decoherence of the collective spin wave due to thermal
atomic motion~\cite{mitchell2000dynamics}. This is observable consistently in independent measurements for both Stokes and anti-Stokes heralding cases. This is complemented by the increase of the probability of heralding with noise photons for high optical powers. The increase in $P_1$ with temperature can be analogously mostly attributed to the enhancement of directionality of the emission in the phase-matched direction. We note that in the presented scheme of frequency degenerate and counter-propagating pump-coupling laser beams, the unavoidable noise contributions in both Stokes and anti-Stokes fields is at least 50~\% of the detected light, because the detectors can't distinguish between equally probable scattering from the two beams with opposite wave vectors. This necessarily implies the contribution of events where the detected fields have been scattered at a close to $\pi$ angle geometry, i.e. backward, which in the case of warm atoms leads to a loss of the two-photon correlation even for large coupling beam powers~\cite{mitchell2000dynamics}. The estimated corresponding limit on the achievable two-photon coupling efficiency  in the anti-Stokes heralding configuration results to $\eta_{\rm 2ph}^{\rm max}\approx \eta_{\rm opt, S}\times 0.5 = 0.17$, where $\eta_{\rm opt, S}=0.34$ corresponds to independent conservative estimation of optical transmission and detection efficiency in the Stokes channel.

The situation is very different for the reversed configuration, where Stokes field provides the heralding events and anti-Stokes is the corresponding signal. The position of the interaction area was iteratively optimized to $d=-1.9$~cm. As anticipated, the two-photon coupling efficiency determining the single photon probability $P_1$ is in the case of near-resonant anti-Stokes emission significantly suppressed by the absorption losses. The severity of this effect is obvious from the observable dependence on the basic source parameters shown in the Fig.~ ~\ref{fig:powerTempS}-b). Although the functional behaviour is analogous to that observed in the Stokes signal field, more than a threefold decrease of the two-photon coupling efficiency and the corresponding $P_1$ makes the observability of the QNG barely feasible.

\section*{Characterizations of nonclassicality}

The nonclassical character of the heralded Stokes field was evaluated using a two different parameterizations with complementary phenomenological features. They employ measured time-tagged photon detection events from the Hanbury Brown and Twiss (HBT) setup with two single-photon avalanche photodiodes (SPADs). The time-tagged photo-counts are processed with time bin widths $T_{\rm bin}=5.67$~ns.

\begin{figure}[t!]
\begin{center}
\includegraphics[width=1.\columnwidth]{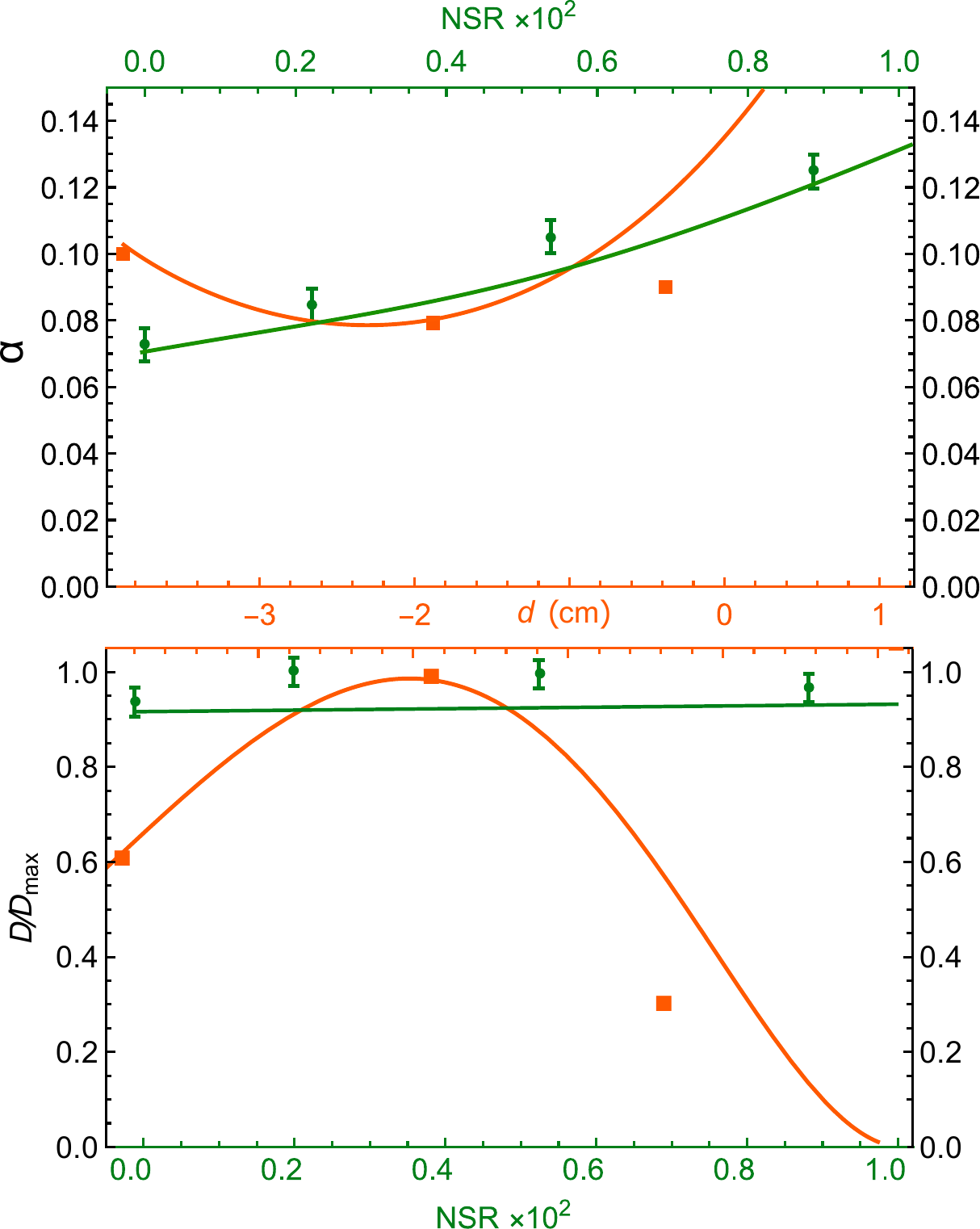}
\caption{Experimental characterizations of the dependence of observable nonclassical statistical character of the heralded Stokes field on position $d$ and on the amount of added thermal noise. The evaluated $\alpha$ and $D$ parameters are shown in graphs a) and b), respectively. Please note, that the point at $d=1.1$~cm gives $\alpha=0.4\pm 0.1$, which is out of the displayed scale.}
\label{fig:dAndAlpha}
\end{center}
\end{figure}

A conventional parameter $\alpha=P_{\rm c}/(P_{\rm s})^2$ is employed for the estimation of the sub-Poissonian features without a direct sensitivity to photon losses, see the Fig.~\ref{fig:dAndAlpha}-a). Here the probability of singles $P_{\rm s}$ has been evaluated as a geometric mean of the probability of single photon detection on the two detectors in the fiber Hanbury-Brown and Twiss setup and $P_{\rm c}$ is the probability of photon coincidence. We additionally prove the nonclassical behaviour of the generated Stokes states by evaluation of the positivity of the witness parameter $D=P_{0}-\sqrt{P_{00}}$, where $P_0$ and $P_{00}$ are the probability of no photon detection event in the given time-bin on a particular SPAD and the probability of no detection on both SPADs, respectively~\cite{lachman2016nonclassical}. Crucially, value of $D$ is insensitive to Poissonian noise while, contrary to the $\alpha$, it is directly dependent on optical losses. Although the probabilities of photon non-detection rates $P_0, P_{00}$ and photon number probabilities $P_1, P_2$ are not independent and parameters $\alpha, D$ can be evaluated one from the other, trajectories in the two representations can directly reveal different aspects of the observed nonclassicality. Fig.~\ref{fig:dAndAlpha}-b) shows the evaluated parameter $D$ normalized to its maximal value $D_{\rm max}$  in order to achieve the optimal scale of the observable dependence. While both parameters unambiguously point to the nonclassical character for all measured positions $d$ and for all probed noise to signal ratios NSR given by the probability of detection of the added thermal light $P_{\rm th}$, presented functional behaviours give insight into different underlaying phenomena. The dependence of $\alpha$ is clearly monotonously deteriorated by addition of thermal noise, which has no visible impact on the $D$. This is expectable in the presented limit of low thermal noise amplitude, in which Poissonian and thermal noise photon probability distributions are almost indistinguishable. The dependence on the position of the interaction area $d$ results in higher observable significant effect on $D$, which is sensitive also to bare loss of photons in the signal Stokes channel, while $\alpha$ is mostly limited by the SFWM Stokes noise caused by a decrease of the heralding efficiency in the anti-Stokes channel. The simulation of the measured behaviour for different $d$ parameters shown as orange solid curves employs precisely estimated spatial overlap of the interaction region with atoms in the cell in the proximity of the output viewport and absorption losses for the anti-Stokes field upon passing the atoms scaled by the Beer–Lambert law with position $d$. The error bars correspond to a single standard deviation and were estimated from a set of five consecutive measurements. We remind that the ${\rm NSR}=P_{\rm th}/P_1=0.94$~\%, where $P_{\rm th}=N_{\rm th} T_{bin}$.

\end{document}